\documentclass[twocolumn,showpacs,prl,aps,10pt,superscriptaddress]{revtex4-1}
\usepackage{amssymb}
\usepackage{amsmath}
\usepackage[pdftex]{graphicx}
\usepackage{color,comment}

\definecolor{grey}{rgb}{.6,.6,.6}

\newcommand{\PEph}{P_{\rm occ}}

\def    \bse{\begin{subequations}}
\def    \ese{\end{subequations}}

\newcommand\ee[1]{e^{#1}}
\newcommand\acd{\hat{a}^\dagger}
\newcommand\ac{\hat{a}}

\newcommand\PP{\mathcal{K}}

\newcommand\ii{i}

\newcommand\cor[1]{\langle #1 \rangle}

\newcommand\hH{\hat{H}}
\newcommand\hc{\hat{c}}

\newcommand\ket[1]{| #1 \rangle}

\newcommand\beq{\begin{equation}}
\newcommand\eeq{\end{equation}}
\newcommand{\bea}{\begin{eqnarray}}
\newcommand{\eea}{\end{eqnarray}}

\newcommand{\tot}{\rm{tot}}
\newcommand{\occ}{\rm{occ}}

\begin{document}


\title{Photon-assisted tunneling with non-classical light}
\author{J.-R. Souquet}
\affiliation{Laboratoire de Physique des Solides, Universit\'e Paris-Sud, 91405 Orsay, France}
\affiliation{Department of Physics, McGill University, Montr\'eal, QC, Canada}
\author{M.~J. Woolley}
\affiliation{School of Engineering, University of New South Wales, ADFA, Canberra, ACT, 2600, Australia}
\author{J. Gabelli}
\affiliation{Laboratoire de Physique des Solides, Universit\'e Paris-Sud, 91405 Orsay, France}
\author{P. Simon}
\affiliation{Laboratoire de Physique des Solides, Universit\'e Paris-Sud, 91405 Orsay, France}
\author{A. A. Clerk}
\affiliation{Department of Physics, McGill University, Montr\'eal, QC, Canada}
 
 \date{July 23, 2014}

\begin{abstract}
{\bf Among the most exciting recent advances in the field of superconducting quantum circuits is the ability to coherently couple 
microwave photons in low-loss cavities to quantum electronic conductors (e.g.~semiconductor quantum dots or carbon nanotubes).
These hybrid quantum systems hold great promise for quantum information processing applications; even more strikingly, they enable exploration of completely new physical regimes.  Here we study theoretically the new physics emerging when a quantum electronic conductor is exposed to non-classical microwaves (e.g.~squeezed states, Fock states).  We study this interplay in the experimentally-relevant situation where a superconducting microwave cavity is coupled to a conductor in the tunneling regime.  We find the quantum conductor acts as a non-trivial probe of the microwave state; in particular, the emission and absorption of photons by the conductor is characterized by a non-positive definite quasi-probability distribution.  This negativity has a direct influence on the conductance of the conductor.}
\end{abstract}

\maketitle

The physics of a tunnel junction illuminated by a purely classical microwave field has been understood since the 1960's with the classic work of Tien and Gordon \cite{tien-gordon}.  This situation is equivalent to simply having an ac bias voltage across the conductor, and the resulting modification of the current is known as photon-assisted tunneling; it has 
been measured in countless experiments (e.g.~Refs.~\cite{Tucker1985,Kouwenhoven1994,Reulet2008}).  Despite the word "photon" in the effect's name, in this standard formulation there is nothing quantum in the treatment of the applied microwave field.  

In this work, we now consider driving a junction with a quantum microwave field produced in a cavity.  The cavity effectively acts as an ac voltage bias across the conductor (see Fig.~1); by maintaining the cavity in a non-classical state, the junction is exposed to a non-trivial microwave field.  Our goal is to understand how such non-classical microwaves affect
electronic transport. Such
cavity-plus-conductor setups have been realized experimentally, both in experiments using metallic tunnel junctions \cite{holst94,Deblock2010,Hofheinz}, as well as more recent experiments with high-$Q$ microwave cavities coupled to either quantum dots \cite{Frey,Petta} or carbon nanotubes \cite{Kontos}.   
Note that the converse problem of how an electronic conductor can be used to produce non-classical squeezed microwaves was recently studied experimentally
\cite{Reulet2013}.

If the cavity is not driven (i.e.~not coherently populated with photons), the cavity-plus-conductor setup realizes another well-studied quantum transport problem: dynamical Coulomb blockade (DCB) \cite{odintsov,nazarov89,devoret90,girvin90}.  Here, the cavity acts as a structured electromagnetic environment for the junction, one that can absorb (and at non-zero temperature, emit) energy from tunneling electrons.  The standard theory of this effect \cite{devoret90,girvin90,IN} is based on the function $P(E)$, which gives the probability of the environment absorbing an energy $E$ from a tunneling electron.  DCB has been experimentally probed both for non-resonant environments \cite{delsing89,Geerligs1989,cleland90,Pierre2007,pierre} as well as for environments formed by resonators \cite{holst94,Deblock2010,Hofheinz}, with excellent 
theoretical agreement.
In stark contrast to standard DCB, our focus will be on a non-equilibrium environment produced by preparing a cavity in a non-classical state.


\begin{figure}[t]
\begin{center}
\raisebox{0pt}{\includegraphics[width=0.4\textwidth]{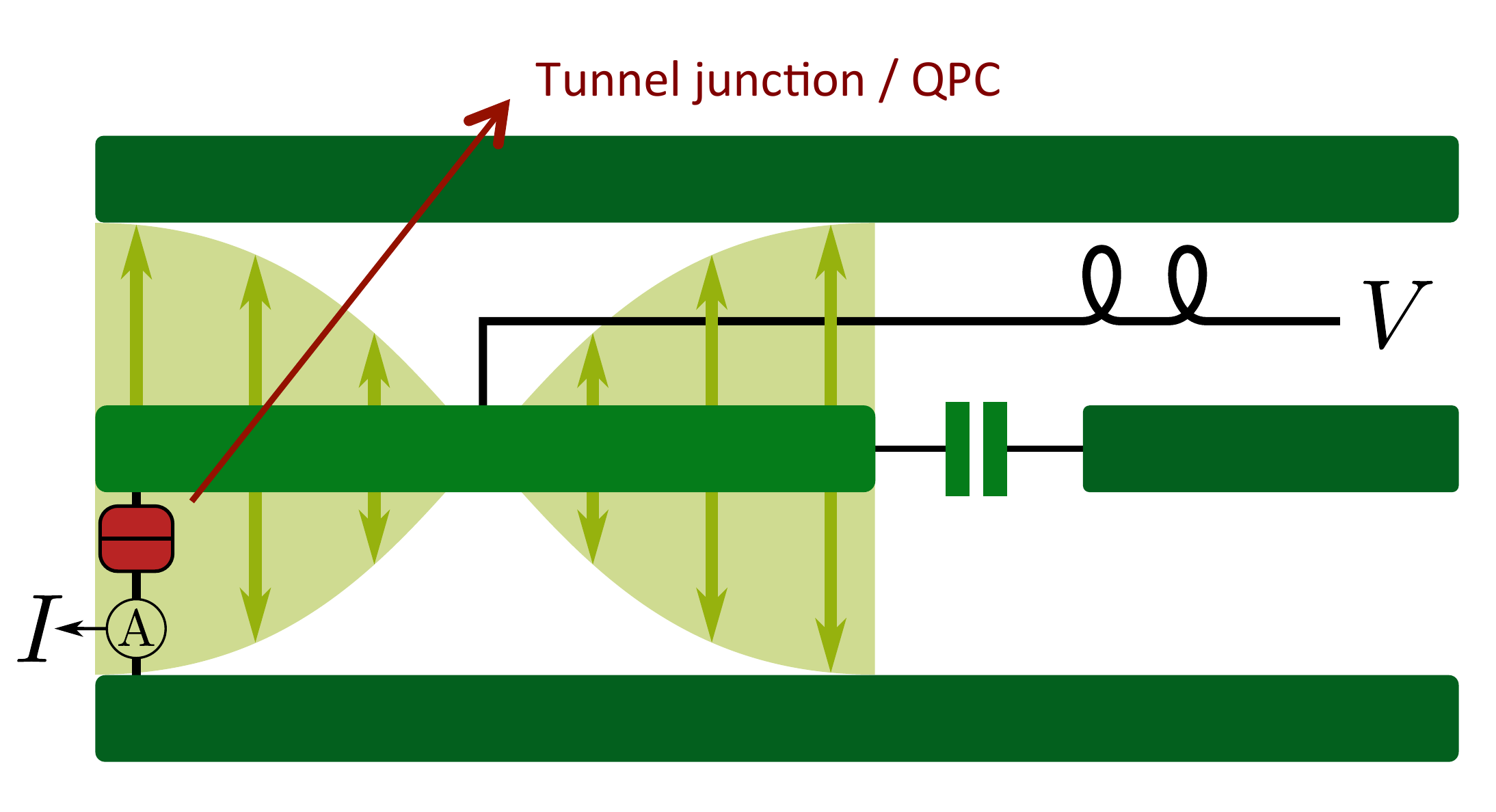}} 
\end{center}
	\caption{ 	\label{fig:schematic} Schematic showing a resonant mode of a half-wavelength coplanar waveguide resonator,
	 with a quantum conductor (tunnel junction or quantum point contact (QPC)) which contacts the centre strip and lower ground plane at a voltage anti-node.
	A dc voltage is also applied to the junction via the centre strip at a voltage node, so as not to induce losses  
	(see, e.g., Refs.~\onlinecite{Rimberg2011,Armour2013,Kycia2013}).  The state of the resonant mode provides a quantum ac voltage across the junction; 
	we are interested in how this influences the dc junction current $I$, and what this current reveals about the quantum voltage.}
\end{figure}

%
 %


{\em Model-- }  As shown in Fig.~1, we consider transport through a voltage-biased tunnel junction (dc bias voltage $V$) which is coupled to the voltage anti-node of a microwave cavity, 
in such a way that the cavity voltage acts as an additional bias voltage across the junction.  We calculate the average current to lowest non-vanishing order in the tunneling strength.  
If the resonator was in thermal equilibrium, we would recover the standard DCB expression \cite{IN}.  We generalize this approach to now allow for an environment (i.e.~the cavity) which is in an 
{\it arbitrary} non-equilibrium, non-stationary state.  In general the average tunnel current is time-dependent, and can be written 
\begin{eqnarray}
	I(t,V) & = &
			e   \sum_{\sigma = \pm} 
				\sigma \int dE \,\, \Gamma(\sigma \cdot eV -  E) P_{\rm tot}(E; t, \sigma).
	\label{eq:IV}
\end{eqnarray}
The two terms here represent (respectively) left-to-right and right-to-left tunneling, 
and $\Gamma(E)$ describes the energy-dependent tunneling rate of the uncoupled junction.  For the usual case of metallic leads, one has
$\Gamma(E) = (  e^2 R_{\rm T})^{-1} E / (1 - \exp(- E / k_B T_{\rm el}) )$, where $T_{\rm el}$ is the lead temperature and $R_{\rm T}$ is the junction resistance 
(see, e.g., Ref.~\cite{NazarovBlanterBook}).  
In this standard case, the current of the uncoupled junction
is purely Ohmic,  $I_0(V) =  V / R_T$.  
The functions $P_{\rm tot}(E; t, \sigma)$ describe energy transfer to/from the electromagnetic environment.  They are given by a causal environment correlation function, evaluated in the absence of tunneling (see SI for derivation):
\begin{eqnarray}
	G_{\rm env}(t, \tau;  \sigma) & = & 
		- (i / \hbar) \theta(\tau) \cor{\ee{  \ii \sigma \hat\varphi(t )}\ee{ - \ii \sigma \hat\varphi(t - \tau)}},  \label{eq:Genv}\\
	P_{\rm tot}(E; t, \sigma) & = & - \frac{1}{\pi} \, \textrm{Im } \int_{-\infty}^\infty \, d \tau \, \ee{\ii E \tau / \hbar} G_{\rm env}(t, \tau;  \sigma).
	\label{eq:pe}
\end{eqnarray}
Here $\hat\varphi= (e / \hbar) \int_{-\infty}^t \hat{U}(t') dt'$ is the phase operator, defined in terms of the (Heisenberg-picture) environment voltage operator $\hat{U}(t)$. 
As shown in the SI, Eqs.~(\ref{eq:IV})-(\ref{eq:pe}) reduce to standard DCB expressions in the usual case of a thermal environment; Eq.~(\ref{eq:pe}) yields a $P_{\rm tot}(E)$ function 
which is positive definite and only depends on $E$.

In our system, we treat the environment as a 
single resonant mode of a cavity, which can be represented as a quantum $LC$ circuit with frequency $\Omega = 1/\sqrt{LC}$. $\hat\varphi$ is thus 
given by one quadrature of the cavity mode annihilation operator $\hat{a}$ \cite{clerk_rmp},
\begin{equation}
	\hat \varphi(t)=- i \sqrt{\rho}\left[ \hat{a}(t) - \hat{a}^\dagger(t) \right],\label{eq:varphi}
\end{equation}
with $\rho=\pi Z_{\rm cav} / R_{\rm K}$ parameterizing the strength of zero-point voltage fluctuations in the cavity ($Z_{\rm cav}=\sqrt{L/C}$, 
$R_{\rm K} = h / e^2$ the resistance quantum).  
Note that Ref.~\onlinecite{Portier2014} recently achieved such a setup with $\rho = 0.3$; higher values should be achievable in the near term.

{\em Closed cavity--}
Our focus will be on situations where the cavity is maintained in some interesting non-vacuum state, either by continuous driving, or via reservoir-engineering
techniques \cite{Zoller1996} which have been used in several recent circuit QED experiments \cite{Murch2012,Devoret2013}.  In either case, this involves coupling the cavity to an external dissipative channel; this gives the cavity
a finite damping rate $\kappa$.  The simplest situation is where this coupling is strong enough to maintain the cavity in the desired state irrespective of the junction current, but still weak enough that
it does not 
appreciably modify the cavity dynamics.  We start by analyzing this situation, meaning that we can neglect the effects of $\kappa$ in calculating $P_{\rm tot}(E;t, \sigma)$; non-zero $\kappa$ will be 
addressed in the next section.  Note that the backaction of the junction on the cavity is formally a higher-order-in-tunneling effect, and thus can be neglected for a sufficiently large tunnel resistance $R_{\rm T}$.  
As discussed
in the SI, if $\rho \simeq 1$, one needs $R_{\rm T} / R_{\rm K} \gtrsim \Omega / \kappa$.

In general, one finds that $P_{\rm tot}(E; t, \sigma)$ and hence the average current oscillates as a function of $t$.  We will focus on the dc current, and thus average over $t$.  The resulting 
$P_{\rm tot}(E)$ function is then only a function of $E$. 
In the $\kappa \rightarrow 0$ limit, the energy of a cavity photon is precisely $\hbar \Omega$, and hence $P_{\rm tot}(E)$ has the form
\begin{equation}
	P_{\rm tot}(E) = \sum_{k=-\infty}^{+\infty} p_{\rm tot}[k] \, \delta(E - k \hbar \Omega).
	\label{eq:PEclosed}
\end{equation}
For a simple Ohmic tunnel junction, the differential dc conductance $dI / dV$ will then exhibit a series of steps as a function of dc voltage $V$, as different photon-assisted processes become energetically allowed.
As discussed in the Methods section, by measuring $dI / dV$ and the (symmetrized) finite-frequency junction current noise $\bar{S}_{I}[\omega,V]$, one can directly extract the weights 
$p_{\rm tot}[k]$.

Without dissipation, the cavity evolves freely, and we can calculate $P_{\rm tot}(E)$ for an arbitrary cavity state $\hat{\rho}_{\rm cav}$.  It can be written as the convolution of two normalized distributions,
\begin{eqnarray}
	P_{\rm tot}(E) = \int dE' \, P_0(E-E') \PEph(E').
	\label{eq:PEConvolution}
\end{eqnarray}
$P_0(E)$ describes the absorption of energy by a ground-state cavity, and only has weight for $E \geq 0$.  In contrast, $\PEph(E)$ is a quasi-probability distribution which describes the additional emission and absorption processes possible when the cavity is occupied with photons.   If the cavity were in its ground state, we would simply have $\PEph(E) = \delta(E)$
and $P_{\rm tot}(E) = P_0(E)$.  
$P_0(E)$ is a Poisson distribution with mean $\rho$  \cite{devoret90,IN}:
\begin{equation}\label{eq:P0closed}
	P_{0}(E) =  \displaystyle \sum_{k = 0 }^{+\infty} \ee{-\rho}\dfrac{\rho^k}{k!} \delta(E-k \hbar \Omega) \equiv  \displaystyle \sum_{k \geq 0 } 
		p_0[k] \,  \delta(E-k \hbar \Omega).
\end{equation}

The function $\PEph(E)$ that we introduce captures the novel physics we are after.  For an arbitrary cavity state
$\hat{\rho}_{\rm cav}$, it is directly related to the Glauber-Sudarshan $P$-function $\PP(\alpha)$ which represents $\hat{\rho}_{\rm cav}$ via a quasi-probability distribution in phase space.  Recall that $\PP(\alpha)$ is defined via
\cite{GerryKnight}
\begin{equation}
	\hat{\rho}_{\rm cav} = \int d^2 \alpha \, \PP(\alpha) |\alpha \rangle \langle \alpha |,
	\label{eq:WFunction}
\end{equation}
where $| \alpha \rangle$ denotes a cavity coherent state with complex amplitude $\alpha$.  $\PP(\alpha)$ expresses $\hat{\rho}_{\rm cav}$ as an incoherent mixture of coherent states.   

For $\kappa \rightarrow 0$, we find that $P_{\rm occ}(E)$ also reduces to a discrete distribution,
\beq
	\label{eq:PEweights}
	\PEph(E) = \sum_{k=-\infty}^{+ \infty} p_{\rm occ}[k] \, \delta(E - k \hbar \Omega),
\eeq	
with weights directly determined by the Glauber-Sudarshan $P$ function:
\begin{equation}
	p_{\rm occ}[k] = \int d^2 \alpha  \,  \PP[\alpha] \left[ J_k\left(2\sqrt{\rho}|\alpha|\right) \right]^{2}.
	\label{eq:PoccNice}
\end{equation}
Here $J_k$ is a Bessel function. 

If $\PEph(E)$ is positive definite, Eq.~(\ref{eq:PEConvolution}) implies that we can interpret the energy $E$ absorbed by the cavity in a tunnel event as the sum of two independent stochastic quantities:  
an amount associated with vacuum fluctuations (as described by $P_0$), and an amount associated with the population of the cavity (as described by $\PEph$).
While $P_{\rm tot}(E)$ must always be positive definite (see SI), this is not necessarily true of $P_{\rm occ}(E)$:  it can become negative for non-classical cavity states, 
i.e.~states whose
phase-space distribution $\PP[\alpha]$ either 
fails to be positive definite or is highly singular \cite{GerryKnight}.  Negativity in $P_{\rm occ}(E)$ will thus be a direct sign of non-classical light.

For further intuition into Eq.~(\ref{eq:PoccNice}), consider the simple case where the cavity is in a coherent state with amplitude $\langle \hat{a}(0) \rangle = \alpha_0$.
In this case 
$\PP[\alpha] = \delta^{(2)}(\alpha - \alpha_0)$, and
\beq
\label{eq:TG}
	p_{\rm occ}[k] =  J_k^2\left(2 \sqrt{\rho} |\alpha_0|\right).
\eeq
$p_{\rm occ}[k]$
is precisely the weight for an $k$-photon process in the standard Tien-Gordon theory for a purely classical ac voltage $V_{\rm ac}(t) \propto |\alpha_0|$  \cite{tien-gordon}.
Thus, Eq.~(\ref{eq:PoccNice}) demonstrates that for a general state,  
$p_{\rm occ}[k]$ is a superposition of Tien-Gordon distributions for different amplitudes, with each term 
weighted by the Glauber-Sudarshan P-function $\PP[\alpha]$.  

Returning to the coherent state case, we see from Eq.~(\ref{eq:PEConvolution}) that the full distribution 
$P_{\rm tot}(E)$ involves convolving the Tien-Gordon distribution with the zero-temperature absorption processes of the cavity.  
This thus generalizes Tien-Gordon theory to include the contribution of cavity vacuum noise.  
Note that a purely classical ac voltage does not modify dc $I$-$V$ characteristic of a conventional tunnel junction, due to the lack of any rectification (i.e.~such a junction
has a purely linear $I$-$V$ characteristic).  This is however no longer true when we include zero-point
fluctuations of the field:  now, the dc $I$-$V$ characteristic of the junction is indeed modified by the presence of the ac voltage.  This behaviour is demonstrated in 
Fig.~\ref{fig:CoherentDrive}.

\begin{figure*}[htpb]
	\begin{center}
		\includegraphics[width=0.4\textwidth]{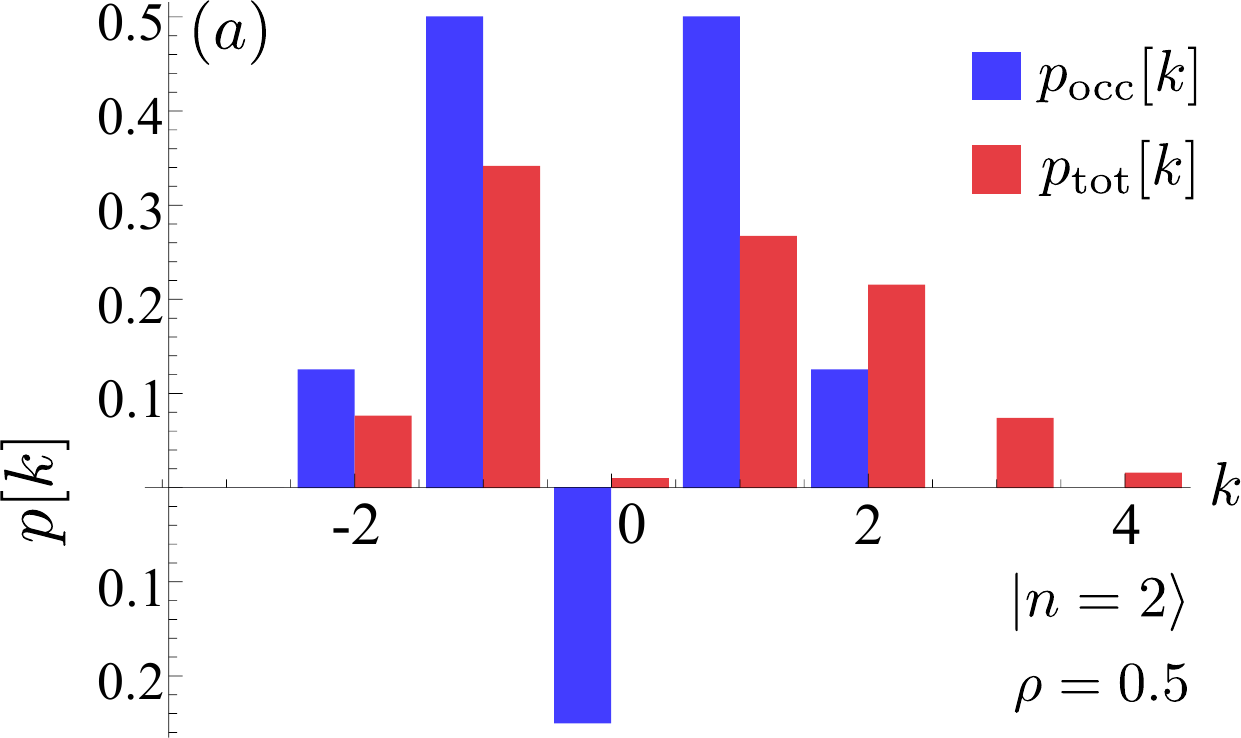}
		\hspace{1cm}
		\includegraphics[width=0.4\textwidth]{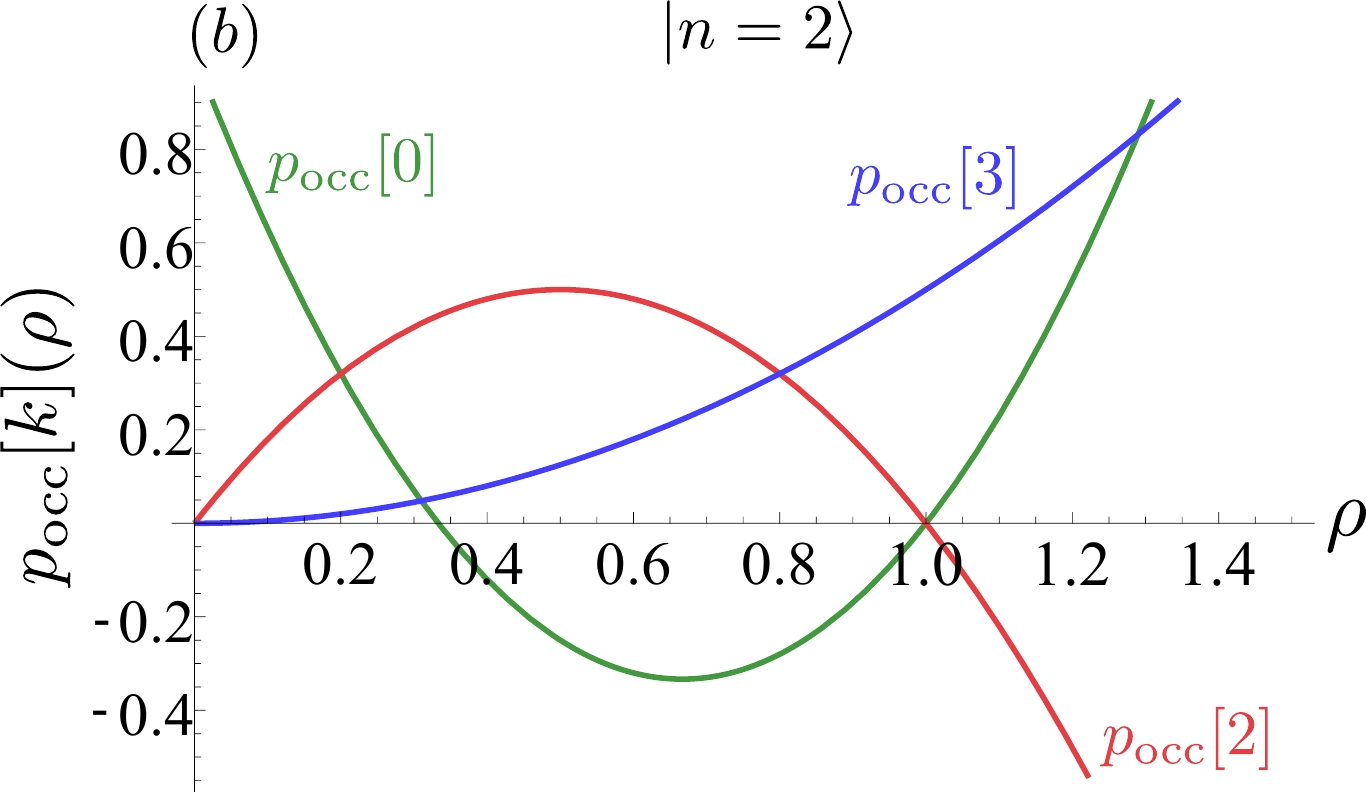}
		\caption{
		\label{fig:FockStateProbs}
		(a)  Probability distributions describing photon emission and absorption by a cavity initially prepared in the $n=2$ Fock state, in the absence of cavity damping,
		and for a dimensionless cavity impedance $\rho \equiv \pi Z_{\rm cav} / R_K = 0.5$.
		The quasi-probabilities $p_{\rm occ}[k]$ characterize the additional photon emission / absorption processes possible due to populating the cavity with photons, whereas the probabilities 
		 $p_{\rm tot}[k]$ also include
		the absorption events associated with vacuum noise.  While $p_{\rm tot}[k]$ must always be positive definite, $p_{\rm occ}[k]$ can fail to be positive for non-classical cavity states.  Here, we see that the weight 
		$p_{\rm occ}[k=0] \leq 0.$
		(b) Behaviour of the quasi-probabilities $p_{\rm occ}[k]$  for a closed cavity in the $n=2$ Fock state,
		as a function of $\rho$ (which characterizes the strength of cavity zero-point voltage fluctuations seen by the conductor).  Negativity requires sufficiently large $\rho$, though the minimum required $\rho$
		decreases with increasing $n$. }
	\end{center}
\end{figure*}

{\it Fock state-- }
Consider now the case where the cavity is stabilized in a Fock state $| n \rangle$; this has been achieved recently via reservoir engineering protocols in circuit QED
\cite{YaleFockState}.  
For the simple case $n=1$, one finds $p_{\rm occ}[k] = 0$ unless $k=0,\pm1$, in which case:
\begin{eqnarray}
	p_{\rm occ}[0] = 1- 2 \rho  , \,\,\, p_{\rm occ}[\pm 1] =   \rho.
\end{eqnarray}
 $\PEph(E)$ for this state describes the possibility to emit or absorb 0 or $1$ photons due to the non-zero cavity population.  
 The quasi-probability for the $0$-photon process however becomes negative for
 $\rho > 1/2$.  Similar negativity is found for other Fock states (see SI and Fig.~\ref{fig:FockStateProbs}); the larger the value of $n$, the smaller the value of $\rho$ needed to see negativity.
 As discussed, this negativity is a direct consequence of the non-classical nature of the cavity state.

\begin{figure*}[htpb]
	\begin{center}
		\includegraphics[width=0.4\textwidth]{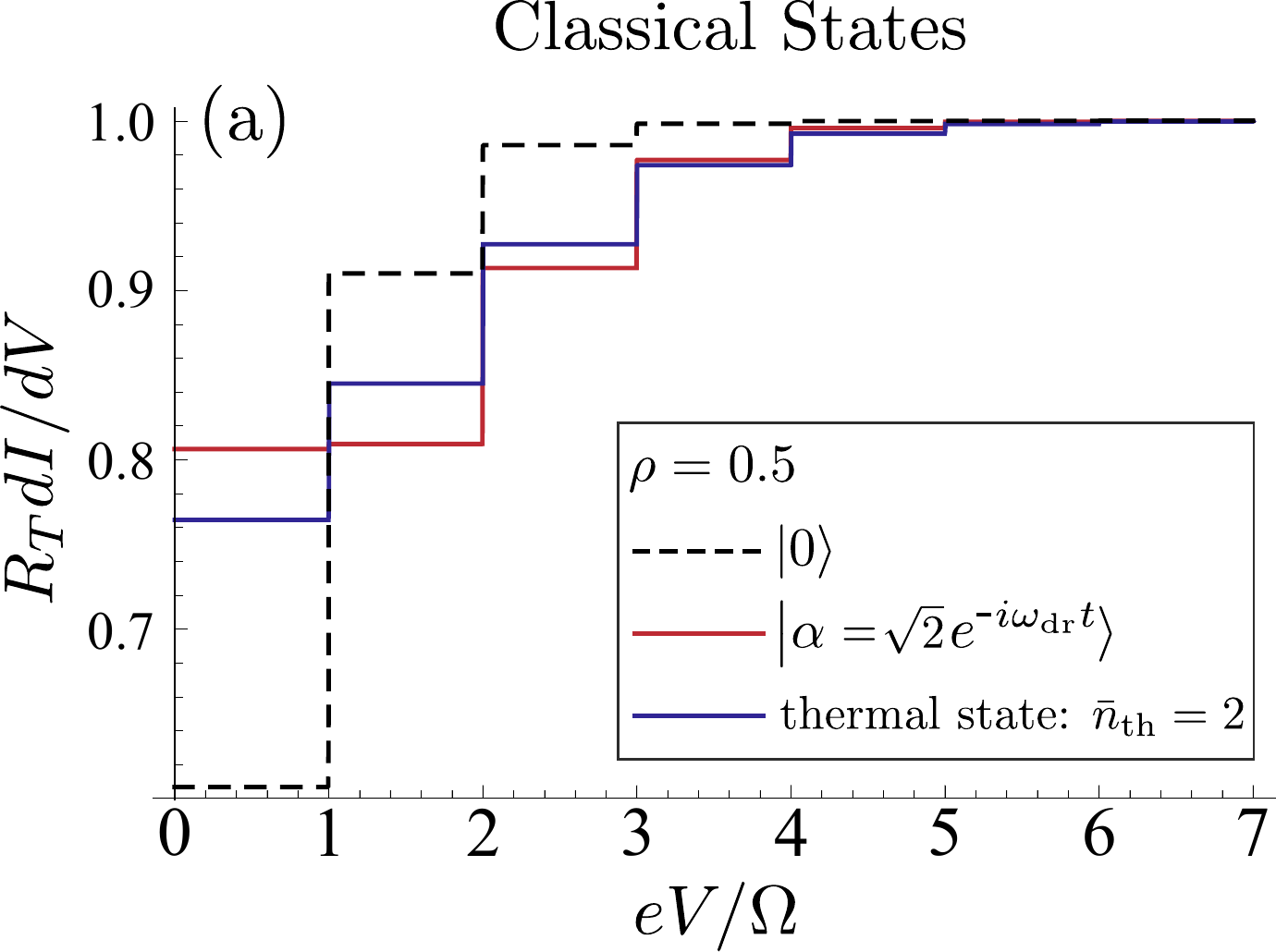}
		\hspace{1cm}
		\includegraphics[width=0.4\textwidth]{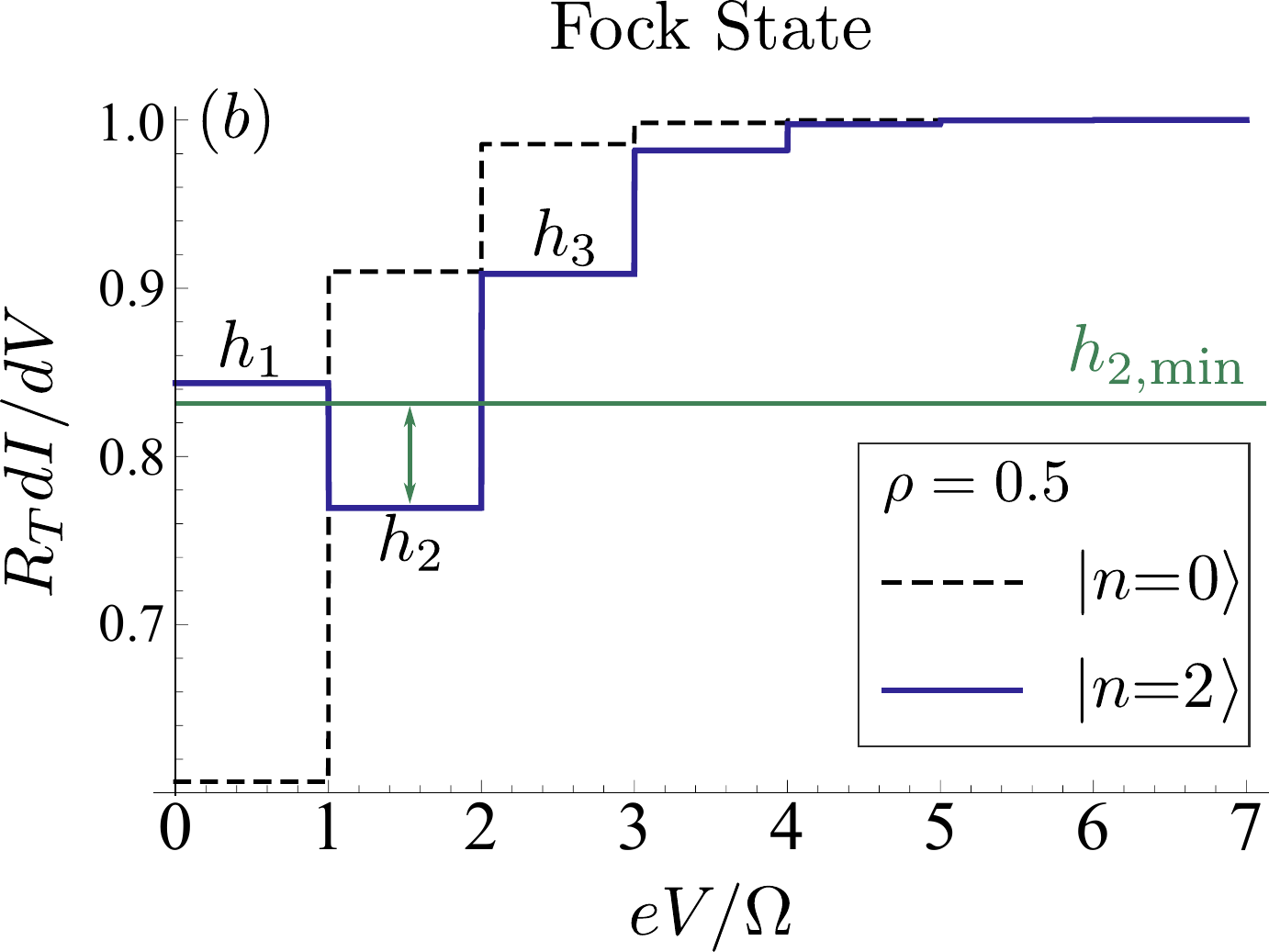}
		\caption{
			\label{fig:FockStateConductance}
		(a)
		Differential conductance $d I / dV$ versus dc bias voltage $V$ for a tunnel junction coupled to a cavity having dimensionless impedance $\rho = 0.5$.
		We assume the cavity is initially prepared in some specific state, 		
		and neglect cavity dissipation for simplicity; we also take the limit of a negligible electron temperature, $T_{\rm el} \ll \hbar \Omega / k_B$.  
		The dashed curve is for a ground state cavity, the solid blue curve for a thermal state with an average photon number $\bar{n}_{\rm th} = 2$,
		and the red curve for a coherent state with average photon number $| \alpha|^2  = 2$. For a thermal cavity state, the conductance plateaus are always
		monotonically increasing in height with $V$. 
		(b)  Same as (a), but now the solid curve corresponds to a cavity prepared in the Fock state $|n =2 \rangle$.  The striking
		signature of a non-classical state here is the strongly non-monotonic dependence of the first few conductance plateaus on voltage; in particular,
		the height of the second plateau ($h_2$) is smaller than the first ($h_1$).  This is in sharp contrast to the classical states shown in (a), 
		states which all have an identical average 
		cavity photon number.  
		As discussed in the SI, if one assumes the distribution $\PEph(E)$ describing the cavity is positive-definite, then one can rigorously bound how
		small $h_2$ can be relative to the average height of the 1st and 3rd plateaus.  This bound is shown as the green horizontal line in the figure.  The conductance
		clearly violates this bound, and thus provides direct (and experimentally-accessible) evidence for the negativity in $\PEph(E)$.
		Similar violations are possible with other choices of Fock state; higher $n$ Fock states allow violations at even smaller values of $\rho$ (see SI).
		}
	\end{center}
\end{figure*}

\begin{figure*}[htpb]
	\begin{center}
		\includegraphics[width=0.4\textwidth]{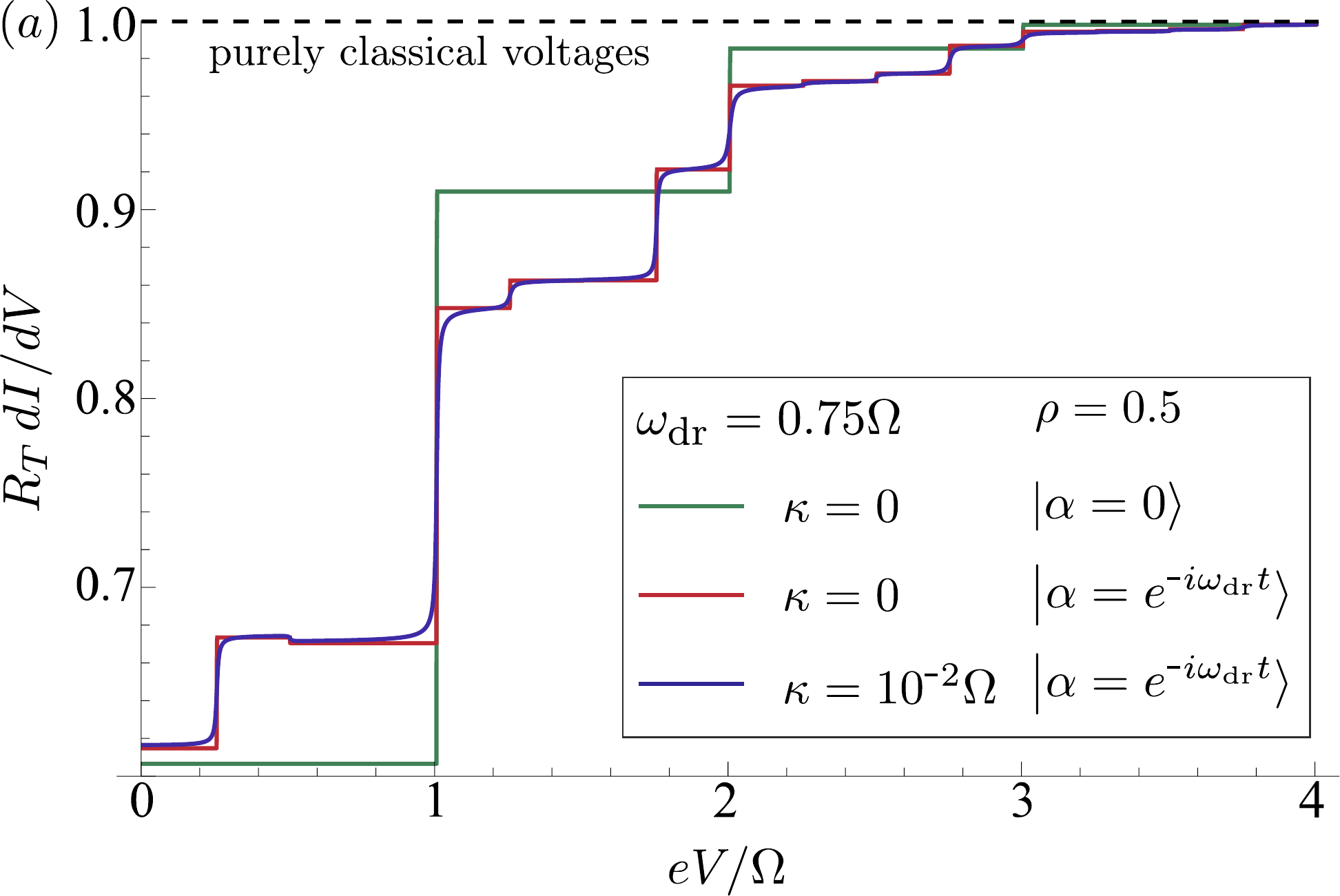}
		\hspace{1cm}
		\includegraphics[width=0.4\textwidth]{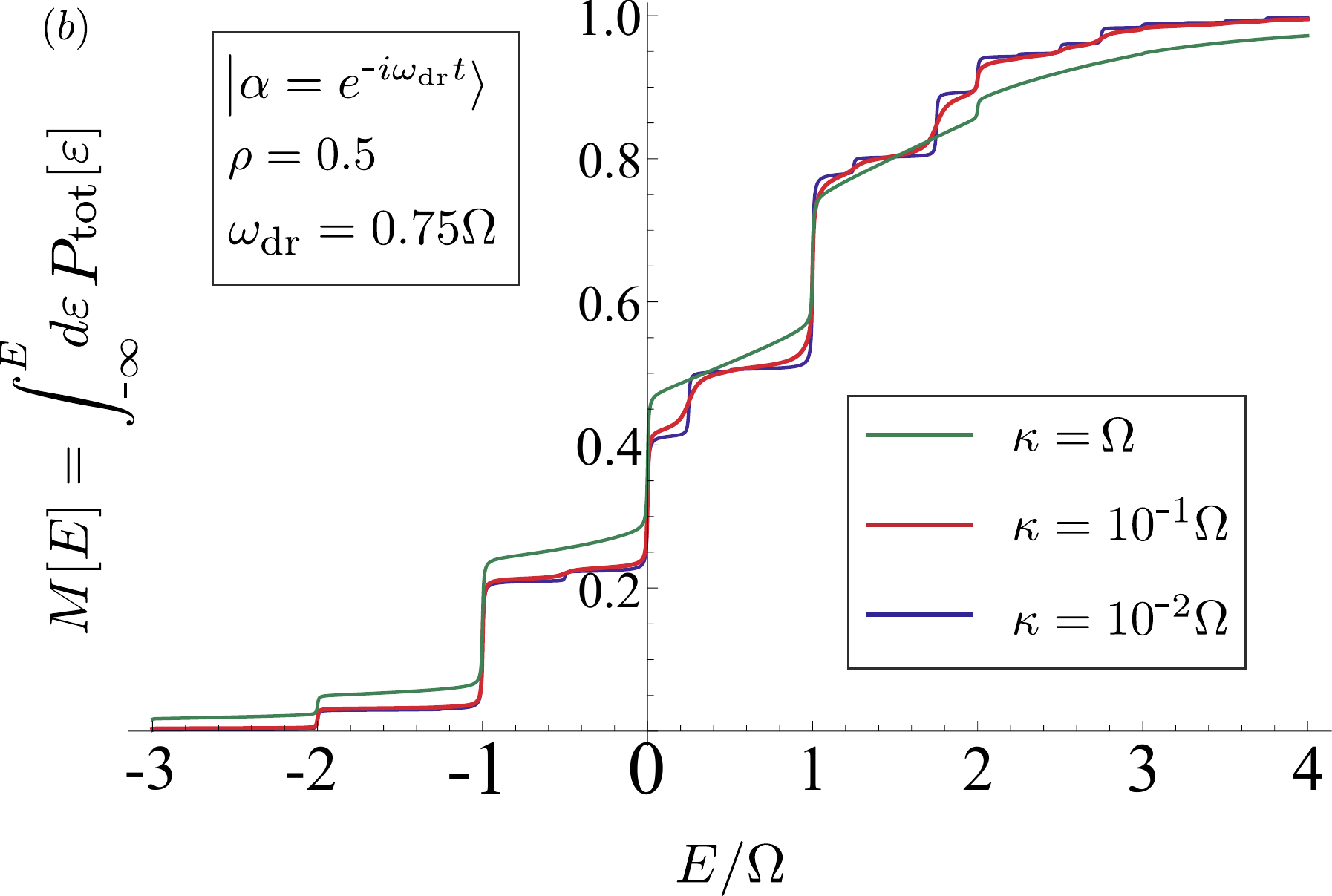}
		\caption{
			\label{fig:CoherentDrive}
		(a) Differential conductance $d I / dV$ versus dc bias voltage $V$ for a tunnel junction coupled to a cavity which is continuously driven into a coherent state having amplitude $| \alpha | = 1$.
		We have taken a drive frequency which is detuned from resonance:  $\omega_{\rm dr} = 0.75 \Omega$.  Results for zero dissipation
		($\kappa \rightarrow 0$) and finite dissipation $\kappa = 0.01 \Omega$ are shown;
		all curves correspond to zero cavity and electron temperature (see SI for finite temperature effects).
		The steps in the conductance now occur at multiples of both the cavity and the drive frequency.  Note that standard photon-assisted tunneling theory 
		(Tien-Gordon theory \cite{tien-gordon}) predicts that $dI/dV = 1 / R_T$ independent of $V$ and the ac voltage.
		Classically, this is due to the linear $I$-$V$ characteristic of a tunnel junction 
		and consequent lack of any rectification.  The behaviour shown here is starkly different, due to the inclusion of zero-point fluctuations.
		(b)  Integrated probability function $P_{\rm tot}(E)$ for the same situation as panel (a).  One again sees steps at multiples of the the cavity resonance frequency and at multiples
		of the drive frequency.  The steps associated with the drive frequency remain sharp even in the presence of cavity dissipation. 
		} 
	\end{center}
\end{figure*}

The negativity in $\PEph(E)$ leads to a distinct signature in the differential dc conductance of the junction (see Fig.~\ref{fig:FockStateConductance}).  
The conductance exhibits regular plateaus as a function of dc voltage.  However, unlike the case of a cavity thermal state, the plateau heights 
associated with a cavity Fock state do not increase
monotonically with voltage.  These surprising decreases in conductance plateau height are inconsistent with
$\PEph(E)$ being positive definite.  As shown in the Methods section, if $\PEph(E)$ were positive, there is a bound on how small the second plateau in $dI/dV$ can be compared to the first
and third plateaus.  This bound is generically violated by the $dI/dV$ obtained with a Fock state in the cavity (e.g. that shown in Fig.~\ref{fig:FockStateConductance}).  
Thus, {\it the differential conductance of the junction provides a direct probe of the non-classical nature of the of the cavity state.} 

Further evidence of the negativity in the Fock state $p_{\rm occ}[k]$ can be seen in the corresponding total emission/absorption probability $p_{\rm tot}[k]$ (which includes the contribution from vacuum noise).  For a cavity maintained in an $n$-photon Fock state, we find (see SI):
\begin{equation}
	p_{{\rm tot},n}[k]= 
		\begin{cases}
      \frac{\ee{-\rho}\rho^kn!}{(k+n)!} \left[L_n^{(k)}(\rho)\right]^2 , & \text{if } k \geq -n, \\
      0	& \text{otherwise}.
		\end{cases}
	\label{eq:FockWeight}
\end{equation}
Here, $L_n^{(k)}$ denotes a generalized Laguerre polynomial.
As expected, if the cavity is maintained in a $n$-photon Fock state, then in a single tunnel event at most $n$ photons can be absorbed.  However, 
for an appropriately chosen $\rho$,  $p_{{\rm tot},n}[-k]$ can be zero for $k \leq n$, while at the same time $p_{{\rm tot},n}[-(k+1)]$ is non-zero.  Such a cancellation would be impossible if $p_{\rm occ}[k]$ is positive definite:  if the probability to absorb $k+1$ photons from the junction is non-zero, then the probability to absorb $k$ photons must also be non-zero.  This is a simple consequence of $p_{\rm tot}[k]$ being the convolution of $p_{\rm occ}[k]$ with a Poisson distribution, $p_{0}[k]$. 

 As discussed in the Methods section, one can directly measure $p_{\rm tot}[m]$ if one measures both the dc conductance of the junction and its finite-frequency current noise.  Using such a measurement to detect the vanishing of 
$p_{{\rm tot},n}[-k]$
for $k \leq n$ would thus also provide direct evidence for the non-classical nature of the cavity state.  If one knows $p_{\rm tot}[m]$, one can also undo the convolution in 
Eq.~(\ref{eq:PEConvolution}) and extract the (possibly negative) quasi-probability distribution $p_{\rm occ}[k]$.  Writing things explicitly, we have:
\begin{equation}
	p_{\rm occ}[k] = 
		e^\rho 
		\sum_{j=0}^{+\infty} 
			\frac{ (-\rho)^j}{j!} p_{\rm tot}[k-j] 
	\label{eq:Deconvolve}
\end{equation}

{\em Cavity driving and dissipation-- }We now consider the case where the cavity is maintained in an interesting state via continuous driving
through an input port, including the non-zero cavity dissipation associated with this port.
Our approach extends easily to such situations if the driving field is Gaussian;
this includes the interesting case of a squeezed vacuum state input.  Letting $\kappa$ denote the damping rate due to the coupling to the transmission line used to drive the cavity, one can use standard input-output theory \cite{clerk_rmp} to derive a Heisenberg-Langevin equation for the cavity field (see Methods).  For Gaussian states, this equation can be solved to obtain the phase-phase correlator and hence 
$P_{\rm tot}(E)$.  

We find that even for a driven, dissipative cavity, $P_{\rm tot}(E)$ can still be written in the general form of Eq.~(\ref{eq:PEConvolution}).  The distribution $P_0(E)$ describes photon absorption by the cavity when it is driven solely by vacuum noise:
\begin{eqnarray}
	P_{0}(E)  & =  &  
		e^{-\rho}\left[
			\delta(E) +
			\sum_{n=1}^{+\infty}  \frac{ (\rho^n / n!) n \hbar \kappa}{ (E-n \hbar \Omega)^2+(\frac{n \hbar \kappa}{2})^2} \right].
\end{eqnarray}
In comparison to Eq.~(\ref{eq:P0closed}), the effects of dissipation are to simply broaden the peaks associated with absorbing $n\geq 1$ photons.  The distribution $\PEph(E)$ again describes additional absorption/emission processes possible when the cavity drive populates the cavity.  

For the coherent state case, we take the cavity to be driven at a frequency $\omega_{\rm dr}$; in this case the average cavity amplitude is $\langle \hat{a} \rangle = \alpha_{0} e^{-i \omega_{\rm dr} t}$.  We find that $\PEph(E)$ is again given by the closed-cavity expression Eq.~(\ref{eq:PEweights})-(\ref{eq:TG}), except that one replaces the cavity frequency $\Omega$ with the drive frequency 
$\omega_{\rm dr}$. In contrast to the vacuum absorption peaks, these processes are not lifetime broadened, and correspond to a photon frequency set by the drive frequency 
$\omega_{\rm dr}$, and not the cavity resonance frequency $\Omega$.  Both these features lead to interesting signatures in the differential conductance; in particular, one sees steps in the conductance corresponding to both relevant photon frequencies (the drive frequency, and the cavity resonance frequency).  This behaviour is demonstrated in 
Fig.~\ref{fig:CoherentDrive}.

{\em Squeezed state-- }
Consider next a cavity which is maintained in a squeezed state, where the variance of one quadrature is reduced below the zero-point value by a factor $e^{-2r}$
($r>0$).
While such a state is Gaussian, it yields a highly singular Glauber-Sudarshan $P$-function, and is thus considered to be non-classical \cite{GerryKnight}.
A squeezed state could be maintained in a superconducting cavity using reservoir engineering techniques \cite{Didier2014}.  Alternatively, one could simply drive the cavity
with squeezed vacuum noise (as produced by a parametric amplifier); this kind of intracavity squeezing has been recently realized in experiment \cite{Murch2013}.  We focus on this situation in what follows. 

Analytic expressions can be obtained for $\PEph(E)$ in the case of a cavity driven by squeezed microwaves, see Eq.~(\ref{eq:S29}) in SI.  One finds that $\PEph(E)$ for $E \simeq \pm \hbar \Omega$ can become negative when $\rho \simeq 1$.  As discussed in the caption of Fig.~\ref{fig:SqueezedState}, this leads to a striking suppression of the peak in $P_{\rm tot}(E)$ near $E = - \hbar \Omega$, which describes the possibility to absorb a single photon.  The weight of this process is suppressed more than would ever be possible if $\PEph(E)$ were positive definite.  Thus, by measuring $P_{\rm tot}(E)$, for a squeezed state, one could directly infer the negativity of $\PEph(E)$. 

As shown in Fig.~\ref{fig:SqueezedState}, this negativity-induced suppression of $P_{\rm tot}(E)$ yields a direct signature in the conductance:  the height of the fourth conductance plateau is higher than would
be possible with any positive definite $P_{\rm occ}(E)$.  In this figure, we also show results including finite cavity dissipation; for small levels of dissipation ($\kappa / \Omega \sim 10^{-3}$) the results are unchanged.   $P_{\rm tot}(E)$ could also be extracted directly if one measures both the differential conductance of the junction and the finite-frequency junction current noise (see Methods).    Note that the finite-frequency current noise measurements for QPC's having $R_T > R_K$ (as we require here) have been performed previously \cite{Glattli2007}.

\begin{figure*}[htpb]
	\begin{center}
		\includegraphics[width=0.4\textwidth]{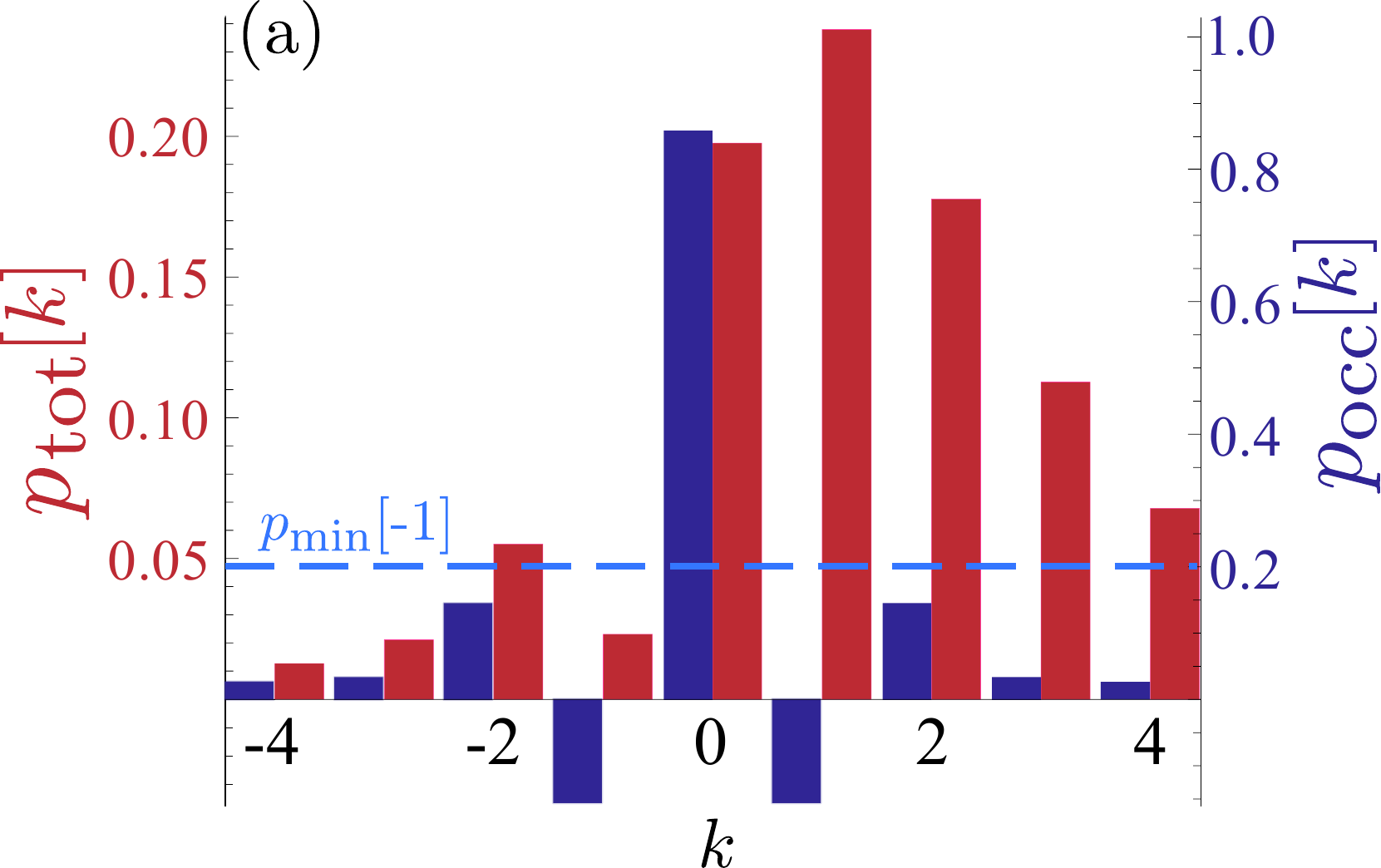}
		\hspace{1cm}
		\includegraphics[width=0.4\textwidth]{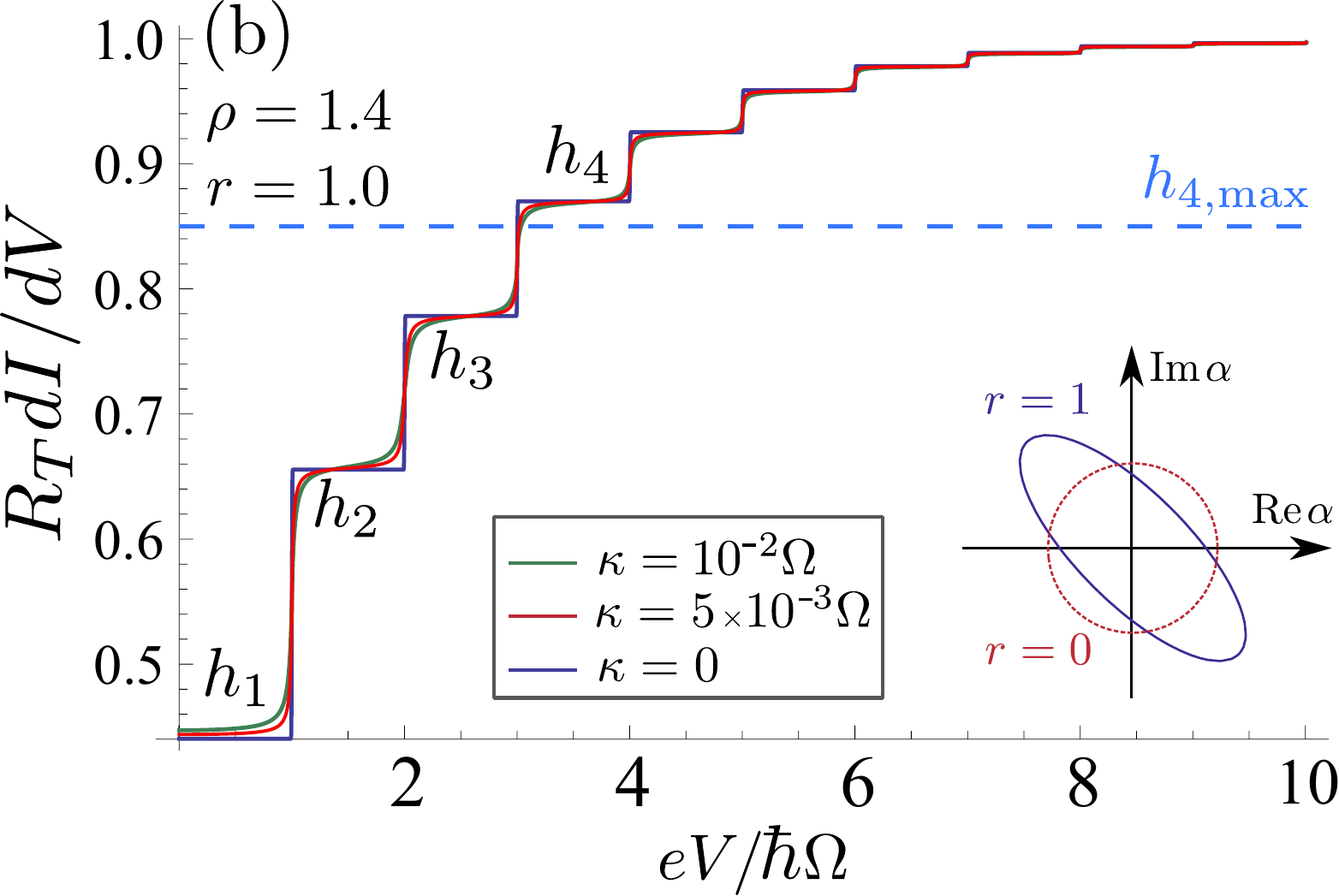}
		\caption{
		\label{fig:SqueezedState}
		(a) Probability distributions $p_{\rm tot}[k]$ and $p_{\rm occ}[k]$ associated with a dissipation-free cavity in a squeezed state vacuum state, with $\rho = 1.4$.
		We take the squeeze parameter to be $r=1$, meaning that the variance of one cavity quadrature is reduced by a factor $1/e^2 \sim 0.14$ below its value in the ground state (see inset of (b)).
		$p_{\rm occ}[k]$ describes the extra photon absorption and emission processes possible when the cavity is occupied with photons, whereas the distribution
		$p_{\rm tot}[k]$ also includes the additional absoprtion processes associated from vacuum fluctuations.  		
		 Similar to a Fock state,
		the squeezed state quasi-probabilities $p_{\rm occ}[k]$ can exhibit negativity, which occurs here most strongly for $k = \pm 1$.  This in turn leads to a 
		strong suppression in the value of the distribution
		$p_{\rm tot}[k]$ at $k=-1$.  If $p_{\rm tot}[k]$ were positive, then $p_{\rm tot}[-1]$ has a minimum possible value $p_{\rm min}[-1]$ (dashed green line); 
		this lower bound is based on the values of $p_{\rm tot}[k]$ at $k=-2, -3$.  
		As clearly shown in the figure, 
		the negativity in $p_{\rm occ}[\pm 1]$ causes a large violation of this bound.
		(b)  Differential conductance for the the same setup in (a); 
		we now however also include the effects of non-zero cavity
		damping $\kappa$. 
		By measuring the heights of the first three conductance plateaus ($h_1$ - $h_3$), one can bound the maximum possible value of the fourth 
		plateau ($h_4$) possible with
		any positive definite $p_{\rm occ}[k]$.  This value is $h_{4,{\rm max}}$, and is indicated with a horizontal line.  We see that the conductance violates this bound,
		and thus provides direct evidence for the negativity of $p_{\rm occ}[k]$.  One can also directly measure the $p_{\rm tot}[k]$ shown in panel (a) by combining the
		conductance measurement shown here with a measurement of the excess current noise $\Delta \bar{S}_{I}[\omega, V] \equiv \bar{S}_{I}[\omega,V]-\bar{S}_{I}[\omega,0]$ (see methods).
		All curves correspond to zero electron and cavity temperatures (see SI for finite temperature effects).
		}
	\end{center}
\end{figure*}

{\em Conclusion-- }We have studied the interplay of non-classical light with electron transport through a tunnel junction, showing that this basic light-matter interaction is naturally
characterized by negative quasi-probabilities for truly quantum states.  This negativity leads to direct signatures in the differential conductance of the conductor, signatures that should be accessible in  state-of-the-art experiments.  Our results can directly be generalized to describe biased Josephson junctions interacting with quantum light; such systems allow even larger values of $\rho$ \cite{IN}.  They also suggest the general potential of using quantum conductors as a powerful tool to characterize, and perhaps control, quantum microwave states in hybrid systems incorporating superconducting microwave cavities and semiconductor electronic devices.


{\bf Methods}
\\
{\em  Closed cavity $P_{\rm occ}(E)$--  }
Using the definition of the Glaubner-Sudarshan $P$ function $\PP(\alpha)$ in Eq.~(\ref{eq:WFunction}) and the fact that $\hat{a}(t) = \hat{a}(0) e^{-i \Omega t}$ for a closed cavity, 
one can explicitly calculate the RHS of Eq.~(\ref{eq:pe}).
Averaging over the observation time $t$ then yields Eq.~(\ref{eq:PoccNice}).  

Alternatively, one can express $P_{\rm occ}(E)$ as ($\hbar=1$)
\bse
\label{eqs:PEFormula}
\begin{eqnarray}
	\PEph(E) & = & 
		\int_{-\infty}^{+\infty} d \tau e^{i E \tau}
			\int_0^{2 \pi / \Omega} \frac{d t }{2 \pi / \Omega}			
			\, \chi[ z(t, \tau)], \\
	z(t, \tau) & = & -2 i \sqrt{\rho} e^{i \Omega t } \sin \Omega \tau /2
\end{eqnarray}
\ese
where the characteristic function $\chi[\lambda]$ is defined as
\begin{equation}\label{eq:chi}
	\chi = \mathrm{Tr } (  \hat\rho_{\rm cav}  \ee{\lambda \acd}  \ee{- \lambda^* \ac}),
	\,\,\,
	\PP[\alpha] = \int d \lambda d \lambda^* \, \chi[\lambda] e^{ \lambda^* \alpha - \lambda \alpha^* }.
\end{equation}

It follows that $\PEph(E) = \PEph(-E)$, regardless of the cavity state (i.e.~there is a perfect symmetry between absorption and emission processes, as one might expect
for purely classical noise \cite{clerk_rmp}).
 $\PEph(E)$ is determined by the normal-ordered expectations
 $\langle : \left( \hat{a}^\dagger \hat{a} \right)^m : \rangle$ of the cavity state; the larger the value 
of $\rho$, the more sensitive one is to higher moments. 

{\em Connecting transport to probabilities-- }For low electron temperatures $T_{\rm el} \ll \hbar \Omega / k_B$, the differential conductance of the junction will 
exhibit sharp steps as a function of $V$, with 
transitions at $eV = m \hbar \Omega$; these steps correspond to turning on and off photon-assisted processes.  We define the normalized height of the $m$th step as   
\begin{equation}
	h_{m}=R_T \left. \frac{dI}{dV}\right|_{V = (m+1/2) \hbar \Omega/e}.
\end{equation}
The height of these plateaus is directly linked to $P_{\rm tot}(E)$.  Consider the simplest case where $T_{\rm el} \rightarrow 0$ and $\kappa \rightarrow 0$ (so that the energy
of a cavity photon is precisely $\hbar \Omega$).  By combining Eqs.~(\ref{eq:IV}) and (\ref{eq:PEclosed}) the normalized first plateau height (i.e.~zero-bias conductance) is
\begin{equation}\label{eq:h1}
	h_1= p_{\tot}[0] + 2 \sum_{k = 1}^{+\infty} p_{\rm{tot}}[-k]  ,
\end{equation}
while the height of subsequent plateaus is
\begin{equation}
	h_{n+1}= h_{n} + p_{\tot}[+n] - p_{\rm{tot}}[-n] .
	\label{eq:hn}
\end{equation}
The behaviour of $dI/dV$ with $V$ allows us to easily extract $ \left(p_{\tot}[+n]-p_{\rm{tot}}[-n]\right)$, the probability difference between an $n$-photon absorption and emission process.

To extract the sum of these probabilities (and hence reconstruct the full distribution $p_{\tot}[n]$), one also needs to measure the finite-frequency current noise of the junction.  We define the (symmetrized) finite-frequency current noise of the junction as
\begin{equation}
	\bar{S}_{I}[\omega, V] \equiv
		\frac{1}{2 T}
		\int^{T/2}_{-T/2} d \bar{t}
		\int dt e^{i \omega t} \left \langle
			\left \{
				\hat{I}(t + \bar{t}),\hat{I}(\bar{t}) 
			\right \}
	\right \rangle,
\end{equation}
where $\hat{I}$ is the junction current operator.
The average over $\bar{t}$ is to pick out the stationary part of the noise (with the averaging time $T \gg 1 / \Omega$).  
This noise spectral density depends both on the drain-source voltage $V$ and the cavity state.  In the tunneling regime, and for $eV < \hbar \Omega$, one finds that
the excess noise $\Delta \bar{S}_{I}[\omega, V] \equiv \bar{S}_{I}[\omega, V] - \bar{S}_{I}[\omega, 0]$ exhibits regular peaks as a function of $\omega$, occurring at $\omega = m  \Omega$
\cite{souquet}.  These noise peaks again correspond to the turning on and off of photon-assisted 
transport processes.  In the low-temperature, low-dissipation case, the heights of these peaks can be directly related to $p_{\rm tot}[n]$ \cite{souquet}
\begin{equation}
	  \Delta  \bar{S}_{I}[\omega = n  \Omega ,V]   = \frac{eV}{R_T} \left( p_{\tot}[+n]+p_{\tot}[-n] \right).
	  \label{eq:DeltaS}
\end{equation}
Thus, measuring both the steps in the differential conductance and the peaks in the frequency-dependent excess current noise allow one to directly extract the probabilities $p_{\rm tot}[n]$.  As mentioned in the main text (c.f.~Eq.~(\ref{eq:Deconvolve})), once one has measured $p_{\rm tot}[m]$ (as described above), one can explicitly extract the values of quasi-probabilities 
$p_{\rm occ}[m]$.


{\em Detecting negative quasi-probability-- }  The distribution $p_{\tot}[k]$ governing photon absorption / emission events is a convolution of 
$p_0[k]$ (absorption due to vacuum noise) and $p_{\occ}[k]$ (absorption and emission due to the presence of photons in the cavity). 
$p_0[k]$ is a Poisson distribution, and is completely determined by the cavity frequency and 
dimensionless impedance $\rho$; $\rho$ could be extracted by, e.g., measuring $dI/dV$ for a ground-state cavity.  This then gives a route for detecting
the negative values of $p_{\occ}[k]$ associated with quantum states.  By using the known behaviour of $p_0[k]$, one can derive general bounds on the differential conductance and excess
noise that must be satisfied {\it for any} positive definite $p_{\rm occ}[k]$.  A violation of such a bound provides direct evidence of negativity in $p_{\rm occ}[k]$, and hence of the non-classical nature 
of the cavity state.

For example, for values of $\rho < \sqrt{3}$, one can derive a minimum possible value for $h_2$ consistent with a positive $p_{\rm occ}$ (see SI):
\begin{equation}
	h_2 >  \frac12 (h_1+h_3) - \frac14 \left(p_0[1] + 2p_0[4]\right) \equiv h_{2,\textrm{min}}.
\label{eq:h2inequality}
\end{equation}
Heuristically, this bound tells us that for a positive $p_{\rm occ}$, the second conductance step cannot be arbitrarily lower than the average height of the first and third steps.
As shown in Fig.~\ref{fig:FockStateConductance}, this inequality is violated if one prepares a $\rho = 0.5$ cavity in a $n=2$ Fock state.  Thus, the differential conductance of the junction gives a direct signature of non-classical behaviour. 

In a similar fashion, one can derive bounds on the behaviour of $p_{\rm tot}[k]$ that are true for any positive-definite $p_{\occ}[k]$; such bounds are in general
even more easily violated by the presence of negativity in $p_{\occ}[k]$.  For example, for $\rho < 2$, one finds that any positive definite $p_{\occ}[k]$ must yield (see SI):
\begin{equation}\label{eq:pmin}
	p_{\textrm{tot}}[n] > 
		\dfrac{p_0[1]}{p_0[0]}
		\left(p_{\textrm{tot}}[n-1] -p_{\textrm{tot}}[n-2]\right) \equiv p_{\rm min}[n].
\end{equation}
For $n=1$, this bound is violated for a cavity with $\rho = 1.4$ prepared in a $r=1$ squeezed vacuum state [see Fig.~\ref{fig:SqueezedState}(a)].  This violation can be detected
experimentally, as $p_{\rm tot}[-1]$ can be extracted from the behaviour of $dI / dV$ and $\Delta S_{I}[\omega, V]$, c.f. Eqs.~(\ref{eq:hn}) and (\ref{eq:DeltaS}).


{\em Heisenberg-Langevin equation-- }  A damped, driven cavity can be described using standard input-output theory \cite{clerk_rmp},
with the cavity equation of motion
\beq
	\dot{\ac}=-i\Omega \ac -\frac{\kappa}{2}\ac-\sqrt{\kappa} {\hat a}_{\rm in}(t).
	\label{eq:InOut}
\eeq
Here, $\hat{a}_{\rm in}(t) = \alpha_{\rm in}(t) + \hat{\xi}(t)$ describes the input field on the cavity: it has an average part $\alpha_{\rm in}(t)$ which describes the classical amplitude of the drive,  and a noise part 
$\hat{\xi}(t)$ which describes both thermal and quantum noise incident on the cavity.  As with standard input-output theory treatments, this noise is taken to be operator-valued Gaussian white noise.  
For a coherent state drive, $\hat{\xi}$ describes vacuum noise, and the average cavity
amplitude is $\langle \hat{a} \rangle = \alpha_{\rm{in}} e^{-i \omega_L t}$.  Squeezed input noise can be simply included in the formalism; it corresponds to anomalous correlators 
$\langle \hat{\xi}(t) \hat{\xi}(t') \rangle$ being non-zero.  

 As the input noise is Gaussian and the cavity has no nonlinearities, the cavity will also be in a Gaussian state.  As a result, the 
phase-phase correlator in Eq.\eqref{eq:pe} is completely determined by two-point correlation functions, and is thus easily found from the solution of Eq.~(\ref{eq:InOut}).

{\bf Acknowledgements}
\\
We thank F. Portier and F. Marquardt for helpful discussions.  MJW acknowledges funding from an ECR Grant of UNSW Canberra..
JG and PS acknowledge partial support from ``Investissements d'Avenir" LabEx PALM (ANR-10-LABX-0039-PALM).    
AC acknowledges support from NSERC.

\bibliographystyle{apsrev4-1}
%

\newpage

\begin{widetext}
\renewcommand{\theequation}{S\arabic{equation}}

\setcounter{equation}{0}

\section{Supplemental information}

\section{$P(E)$ theory for a general non-equilibrium environment}

\subsection{General derivation}

In this section, we present the derivation of Eq.~(\ref{eq:IV}) in the main text, which generalizes the standard $P(E)$ theory for tunnelling in the presence of an
electromagnetic environment to cases where the environment is in an arbitrary time-dependent state.  We consider a tunnel junction between left
and right metallic reservoirs which is voltage biased both by a fixed dc voltage $V$ and by a voltage created by a bosonic electromagnetic environment.
The Hamiltonian is
\beq
	\hH= \hH_{\rm el} + \hH_{\rm env} + \hH_{\rm tun},\eeq
where $\hH_{\rm el}$ and $\hH_{\rm env}$ denote the Hamiltonians of the leads and of the cavity respectively, and $\hH_{\rm tun}$
the tunneling between leads.  Making the usual gauge transformation to include voltages directly in $\hH_{\rm tun}$, we have ($\hbar=1$)
\beq\label{eq:htun}
	\hH_{\rm tun}= w \sum\limits_{k,q} \hc^\dag_{R,k} \hc_{L,q} e^{i e V t} e^{i \hat \varphi(t)} +h.c.\, 
	\equiv \hat{W} e^{i e V t} e^{i \hat \varphi(t)} + h.c.,
\eeq
where $w$ is the tunnel matrix element, $\hc_{\alpha,k}$ is the destruction operator for a single particle state $k$ in lead $\alpha$, and $\varphi(t)$ is the phase operator associated with the environment votage (see Eq.~(\ref{eq:varphi}) of the main text).  The tunnel resistance $R_T$ of the junction is given by $1 / R_T = 
(e^2/h) (2 \pi) w^2 \rho_0^2$, where $\rho_0$ is the lead density of states at the Fermi energy.
As usual, the current operator is given by $\hat{I} = -i  \left( \hat{W} - \hat{W}^\dagger \right)$.  Using standard quantum linear response theory (i.e.~the Kubo formula), the
current at time $t$ to order $w^2$ is given by
\begin{eqnarray}
	\langle \hat{I}(t) \rangle & = & e \left[ \Gamma_{+}(t) - \Gamma_{-}(t) \right], \\
	\Gamma_{\sigma}(t) & = &
		\textrm{Re } \int_{-\infty}^{t - t_0}  d \tau \,
		G_{\rm el}(\tau) G_{\rm env}(t, \tau; \sigma) e^{i \sigma e V \tau}, 
	\label{eq:Iexpression}
\end{eqnarray}
where the relevant electronic and environment Green functions are evaluated in the absence of tunnelling, and are given by
\begin{eqnarray}	
	G_{\rm el}(\tau) & = & -i \langle \hat{W}(\tau) \hat{W}^\dagger(0) \rangle, \\
	G_{\rm env}(t, \tau;  \sigma) & = & 
		-i \theta(\tau) \cor{\ee{  \ii \sigma \hat\varphi(t )}\ee{ - \ii \sigma \hat\varphi(t - \tau)}},
\end{eqnarray}
We have used the fact that the uncoupled electronic system is time translationally invariant (while we have not assumed this about the bosonic environment).  Here, $t_0$ corresponds to the time at which the tunnel Hamiltonian was switched on; we let $t_0 \rightarrow - \infty$.  

Next, note that for free electrons:
\begin{equation}
	G_{\rm el}[\omega] = -i \Gamma[\omega] = -i \frac{1}{ e^2 R_T}  \frac{ \omega} { 1 - \exp(- \omega / k_B T_{\rm el}) },
\end{equation}
Using the convolution theorem to evaluate Eq.~(\ref{eq:Iexpression}), we recover Eq.~(\ref{eq:IV}) of the main text.  

\subsection{Time-independent environment}

For the standard case where the environment is in a time-independent state, environmental correlation functions are time-translation invariant, and hence $G_{\rm env}(t, \tau; \sigma)$ (c.f.~Eq.~(\ref{eq:Genv}) of the main text) becomes independent of the time $t$.  It is then easy to show that Eq.~(\ref{eq:pe}) of the main text reduces to
\begin{eqnarray}
	P_{\rm tot}(E; \sigma) = 
		\int_{-\infty}^{\infty} d \tau e^{i E \tau} \cor{\ee{  \ii \sigma \hat\varphi(\tau )}\ee{ - \ii \sigma \hat\varphi(0)}}.
\end{eqnarray}
Further, $P(E)$ theory is usually applied to situations where the environment is invariant under $\hat\varphi \rightarrow - \hat\varphi$.  In this case, there is no dependence on $\sigma = \pm$, and one recovers the standard formula for the $P(E)$ function as the Fourier transform of the environmental phase-phase correlator.

\subsection{Positivity of time-averaged $P_{\rm tot}(E;t)$:  closed cavity}

For a general time-dependent environment, the function $P_{\rm tot}(E; t,\sigma)$ will be explicitly time-dependent and could take on negative values.  We now focus on the simple case where the environment is a closed cavity, prepared in some arbitrary state.  The cavity Hamiltonian is time independent, and all dependence on $t$ arises from preparing the system in a non-stationary state
(i.e. the cavity density matrix $\hat{\rho}_{\rm cav}$ is not diagonal in the basis of energy eigenstates).  In this case, we write
$P_{\rm tot}(E;t,\sigma)$, using the convolution theorem, in the following form ($\hbar = 1$):
\begin{eqnarray}
	P_{\rm tot}(\omega; t, \sigma) & = & 
	\frac{- \textrm{Im }}{\pi} 
		\int d\omega'  \left[ -i \pi \delta(\omega - \omega') + \frac{1}{\omega - \omega'} \right] \Lambda(\omega',t; \sigma), 
		\label{eq:Causal}\\
	\Lambda(\omega; t, \sigma) & = &
		\int_{-\infty}^{\infty} d \tau 
		\cor{\ee{  \ii \sigma \hat\varphi(t  )}\ee{ - \ii \sigma \hat\varphi(t-\tau)}} e^{i \omega \tau}	
		\label{eq:LambdaDefn}
		\\
		& = & 
		\sum_{i,i',f=0}^{\infty} \langle i | e^{i \sigma \hat{\varphi} } | f \rangle \langle f | e^{-i \sigma \hat{\varphi}} | i' \rangle
		\langle i'| \hat{\rho} |i \rangle
		\delta(E_{f} - E_{i'} - \omega) e^{-i (E_{i'} - E_i) t}. 
		\label{eq:LambdaKernel}
\end{eqnarray}
Here, $|j \rangle$ labels energy eigenstates of the system with corresponding eigenvalues $E_j = \hbar \Omega(j + 1/2)$.

In general, $P_{\rm tot}(\omega; t, \sigma)$ and $\langle \hat{I} (t) \rangle$ will oscillate as a function of $t$ with a period  $2 \pi / \Omega$.  To obtain the dc current, we will average $t$ over one period:
\begin{equation}
	P_{\rm tot}(E) \equiv
		\frac{1}{t_{\rm avg}}
		 \int_{-t_{\rm avg}/2}^{t_{\rm avg/2}} 
		dt'
		P_{\rm tot}(E, t+t'; \sigma), 
\end{equation}
where $t_{\rm avg} = 2 \pi / \Omega$.
The time average kills all terms in Eq.~(\ref{eq:LambdaKernel}) except those where $E_{i} = E_{i'}$.  As the cavity has no degeneracies in its spectrum,  it immediately follows that the only terms in 
$\Lambda(\omega, t; \sigma)$ surviving the time average have $i = i'$, and are thus proportional to matrix elements of the form $| \langle f | e^{-i \sigma \hat{\varphi}} | i \rangle |^2$.  It thus follows that all contributing terms to $\Lambda$ are positive definite, and thus so is $P_{\rm tot}(E)$.  Further, such matrix elements are independent of whether $\sigma = \pm 1$; hence, $P_{\rm tot}$ is independent of $\sigma$.

\subsection{Positivity of time-averaged $P_{\rm tot}(E; t)$:  general case}

In the general case, where the total $P(E)$ function is not periodic, we define the time-averaged $P(E)$ function as ($\hbar = 1$):
\begin{eqnarray}
	\bar{P}_{\rm tot}(E; \sigma) \equiv
		\lim_{T \rightarrow \infty}  \frac{1}{T} \int_{-T/2}^{T/2} d t \, P_{\rm tot}(E; t,\sigma),
		\label{eq:PBarDefn}
\end{eqnarray}
where $P_{\rm tot}(E; t,\sigma)$ is defined in Eq.~(\ref{eq:pe}) of the main text.  From Eq.~(\ref{eq:Causal}), we see that $\bar{P}_{\rm tot}(E; \sigma)$ is necessarily positive definite if the quantity
\begin{eqnarray}
	\bar{\Lambda}(\omega, \sigma) = \lim_{T \rightarrow \infty} \frac{1}{T} \int_{-T/2}^{T/2} dt 
	\Lambda(\omega; t, \sigma),
\end{eqnarray}
is positive definite, where $\Lambda(\omega; t, \sigma)$ is defined in Eq.~(\ref{eq:LambdaDefn}) above. To show this, we first define
\begin{eqnarray}
	\hat{A}[\omega] \equiv
		\frac{1}{\sqrt{T}} \int_{-T/2}^{T/2} dt' \,  \hat{A}(t')  e^{i \omega t'}
\end{eqnarray}
for some arbitrary Heisenberg-picture operator $\hat{A}(t)$.  It immediately follows that:
\begin{eqnarray}
	Q(\omega, T) \equiv \left \langle \hat{A}[\omega] \left( \hat{A}[\omega] \right)^\dagger \right \rangle \geq 0.
\end{eqnarray}
We can express $Q(\omega, T)$ as
\begin{eqnarray}
	Q(\omega, T) = \frac{1}{T} \int_{-T/2}^{T/2} d t \int_{t-T/2}^{t+T/2 } d \tau
		\langle \hat{A}(t) \hat{A}^\dagger(t-\tau ) \rangle e^{i \omega \tau}.
		\label{eq:QEqn}
\end{eqnarray}
Assuming that the correlation function in Eq.~(\ref{eq:QEqn}) has a finite correlation time (i.e.~it decays for sufficiently large $|\tau|$), when taking the limit
$T \rightarrow \infty$ we can safely replace the bounds of the $\tau$ integration by $\pm \infty$.
Making the choice $\hat{A}(t) = \ee{  \ii \sigma \hat\varphi(t )}$, we then have
\begin{eqnarray}
	\bar{\Lambda}(\omega, \sigma) \equiv \lim_{T \rightarrow \infty} Q(\omega, T)  \,\,\, \implies \bar{\Lambda}(\omega, \sigma) \geq 0.
\end{eqnarray}
This proves that $\bar{P}_{\rm tot}(E; \sigma)$ must be positive definite.  

Finally, note that in the cases of interest in the main text, $P_{\rm tot}(E; t, \sigma)$ is a periodic function of $t$.  In this case the infinite-time average over $t$ in Eq.~(\ref{eq:PBarDefn}) is equivalent to averaging $t$ over a single period.

%

\subsection{Transport-induced cavity dissipation}

As discussed in the main text, the backaction of the junction on the cavity does not formally influence the current to lowest order in the tunnelling.  Nonetheless, we can use our approach to estimate the typical size of such effects.  Consider first an undamped cavity, and imagine we have calculated $p_{\rm tot}[k]$ for some given cavity state.  We now want to understand how photon-assisted transitions involving the junction lead to heating (or cooling) of the cavity.  Using a Golden rule approach, the rate of change of the average cavity photon number due to such transitions will be given by:
\begin{eqnarray}
	\frac{d}{dt} \langle \hat{a}^\dagger \hat{a} \rangle \Bigg|_{\rm junc} 
		= \sum_k \sum_{\sigma = \pm} k \, p_{\tot}[k]  \Gamma(\sigma eV - k \Omega)\equiv \mathcal{P}_{\rm em},
\end{eqnarray}
where $\mathcal{P}_{\rm em}$ denotes the power emitted by the junction to the cavity (in units of cavity quanta per unit time).  
Consider the case $T_{\rm el} \rightarrow 0$ and $eV < \hbar \Omega$.  In this case, one finds easily:
\begin{eqnarray}
	\mathcal{P}_{\rm em} = - \frac{R_K}{ \pi R_T} \Omega \sum_{k=1}^{\infty} k^2 p_{\rm tot}[-k] 
	\equiv  - \mathcal{P}_0 \sum_{k=1}^{\infty} k^2 p_{\rm tot}[-k]. 
	\label{eq:PowerIn}
\end{eqnarray}
We see that the scale for these heating/cooling effects is determined by the rate $\mathcal{P}_0$, which as expected becomes weaker the weaker the tunnelling.  
For the states and regimes discussed in the main paper, where average photon numbers are $\simeq 1$ and typical voltages $eV \simeq \hbar \Omega$, the scale of the energy flux $\mathcal{P}_{\rm em}$ will be $\sim \mathcal{P}_0$.  In Figs.~\ref{fig:PowerClassicalStates} and \ref{fig:PowerQuantumStates}, we explicitly show this rate for a variety of different states, demonstrating that $\mathcal{P}_0$ is indeed the relevant scale when the dimensionless cavity impedance $\rho \sim 1$ (i.e.~the regime of interest in the main text).  For $\rho \ll 1$, junction induced heating/cooling will be reduced below $\mathcal{P}_0$ by a further factor of $\rho$.

If we now include cavity damping at a rate $\kappa$, a simple rate equation tells us that the change in the cavity photon number due to junction-induced heating/cooling, $\Delta n_{\rm cav}$, will be approximately
\begin{eqnarray}	
	\left|  \Delta n_{\rm cav}  \right| 
		\sim \frac{\left|  \mathcal{P}_{\rm in} \right|}{\kappa} \sim \frac{\mathcal{P}_0}{\kappa}
\end{eqnarray}
for $\rho \sim 1$.
Insisting that this change be much smaller than a single quantum thus results in the condition:
\begin{eqnarray}	
	\frac{R_K}{R_T} < \frac{\kappa}{\Omega}.
\end{eqnarray}
Thus, if we use a cavity with $\kappa = 10^{-2} \Omega$, the above estimate tells us that the junction resistance needs to be much larger than $\sim 10^2 R_q $.  This could be achieved by using, e.g., a single channel quantum point contact deep in the tunnelling regime.

\begin{figure*}[htpb]
	\begin{center}
		\includegraphics[width=0.4\textwidth]{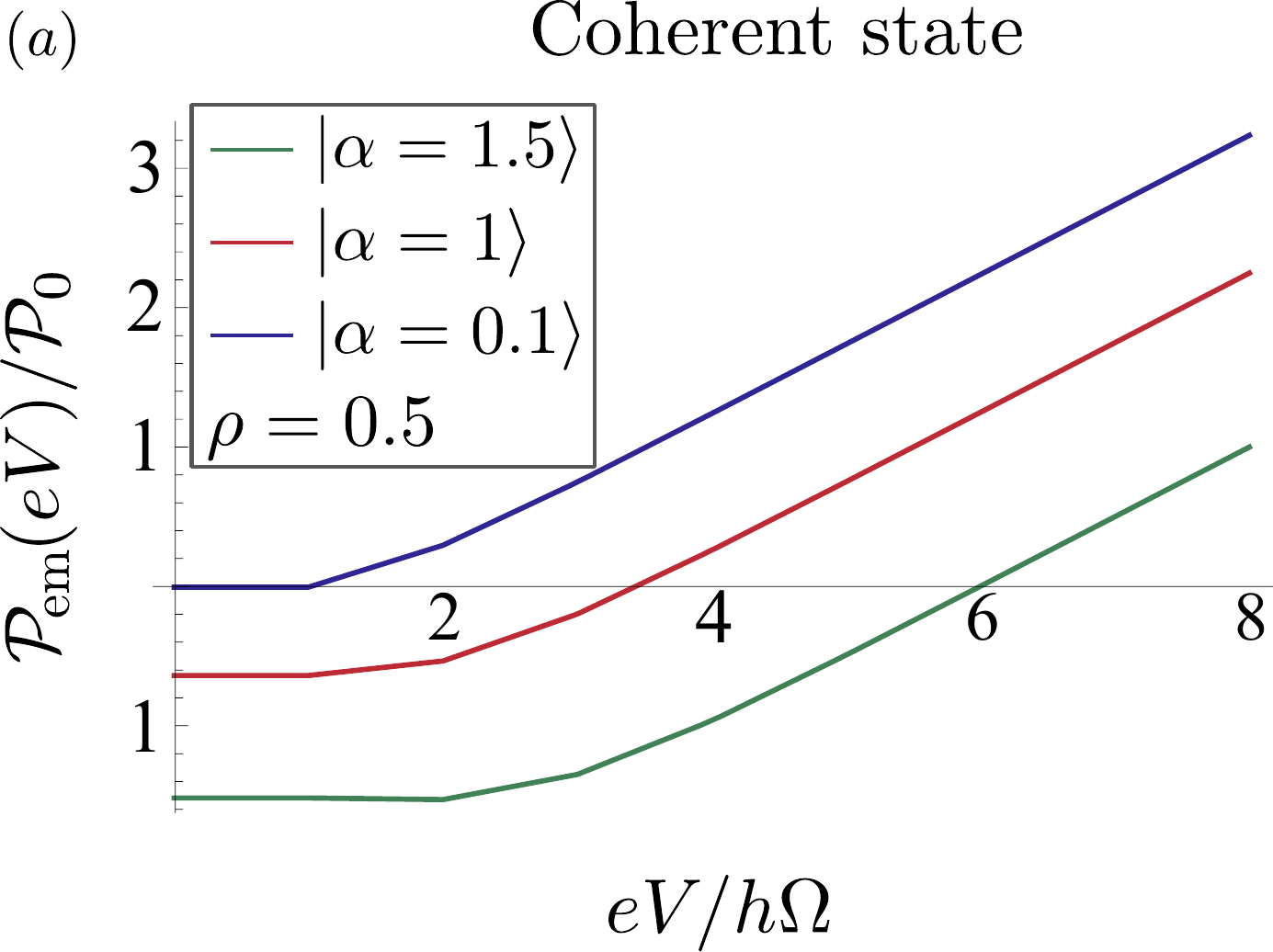}
		\hspace{1cm}
		\includegraphics[width=0.4\textwidth]{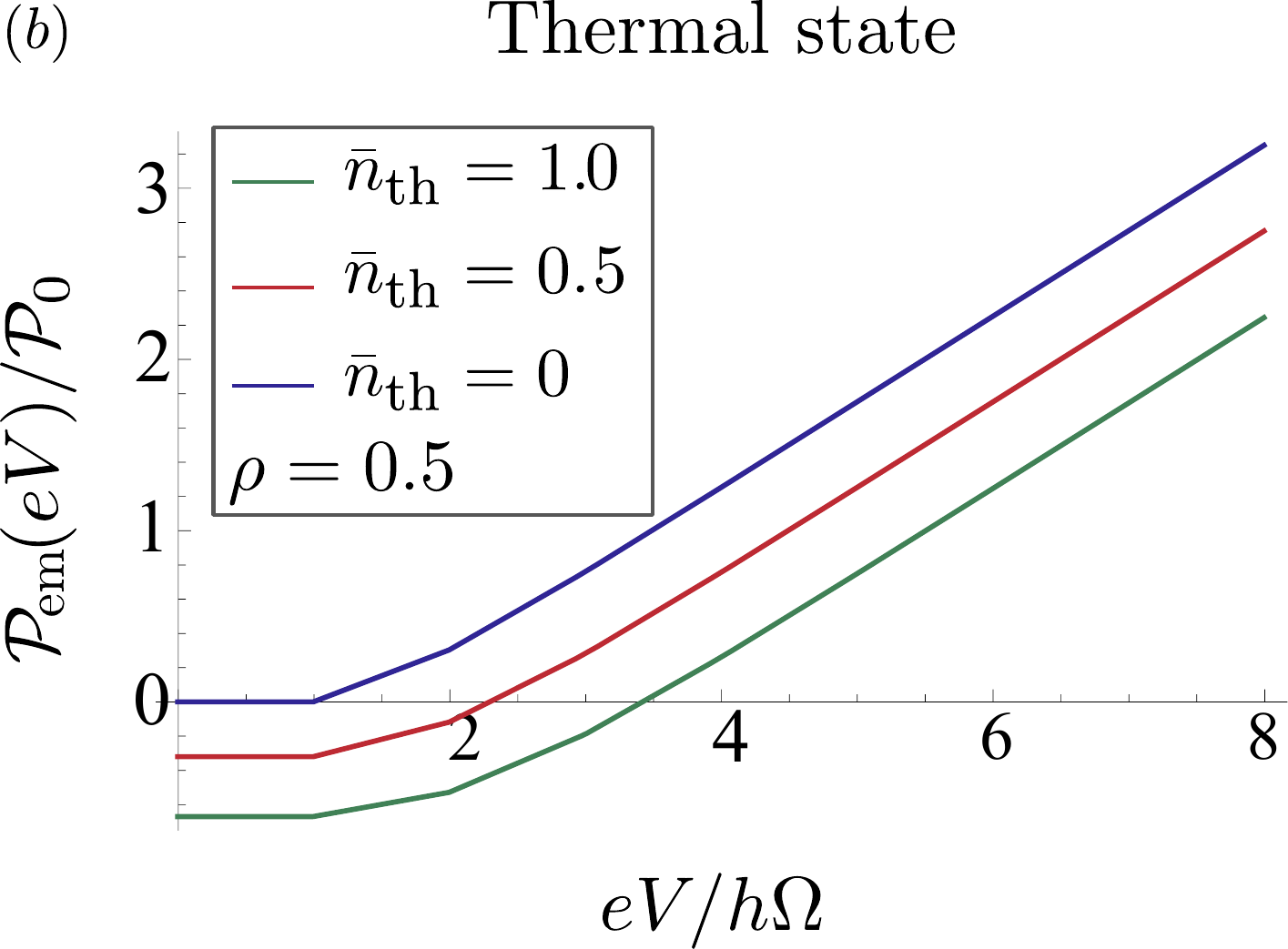}
		\caption{
			\label{fig:PowerClassicalStates}
(a) Power emitted by the junction $\mathcal{P}_{\rm em}$ to the cavity in units of $\mathcal{P}_0$ (c.f.~Eq.~(\ref{eq:PowerIn})), 
as a function of the dc junction bias voltage $V$.
Each curve corresponds to the cavity being in a coherent state with a given amplitude $\alpha$.  For small $V$, the junction acts like a low-temperature bath and acts to cool the cavity (it absorbs energy).  For higher biases, the effective temperature of the junction increases, and there is a net energy flow from the junction to the cavity.
For the range of voltage considered, the emitted power is of order of $\mathcal P_0$.  All curves correspond to zero temperature and zero cavity damping. (b) Same quantity, now for a thermal state in the cavity, 
for different choices of the thermal
photon number $\bar{n}_{\rm th}$.
		}
	\end{center}
\end{figure*}

\begin{figure*}[htpb]
	\begin{center}
		\includegraphics[width=0.4\textwidth]{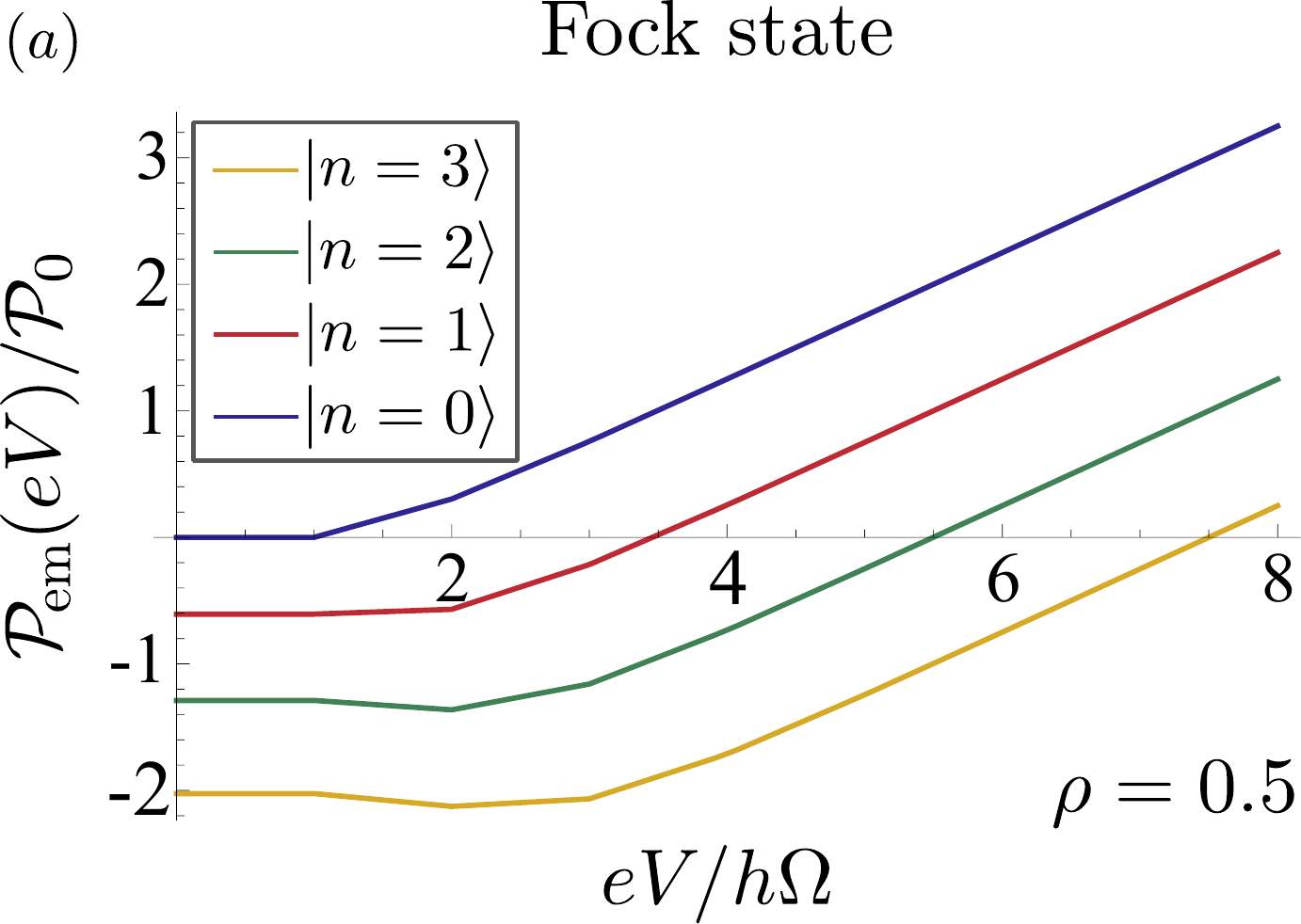}
		\hspace{1cm}
		\includegraphics[width=0.4\textwidth]{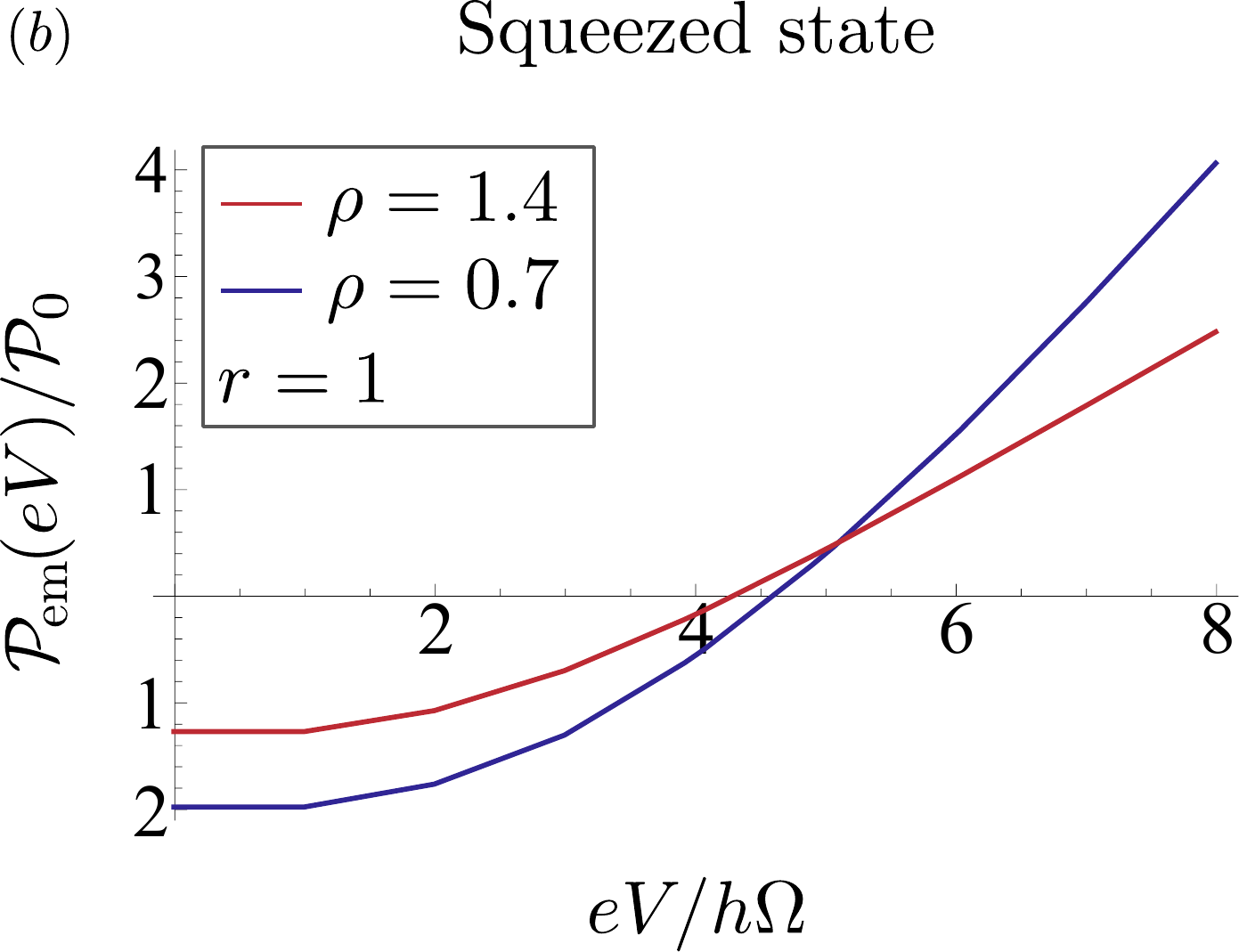}
		\caption{
			\label{fig:PowerQuantumStates}
		Similar to Fig.~\ref{fig:PowerClassicalStates}, but now we take the cavity to be in a non-classical state (as specified in the legend);
		all curves are for zero temperature and for zero cavity damping.
		}
	\end{center}
\end{figure*}

\section{$P_{\rm occ}(E)$ for various closed-cavity states}

In what follows, it will be useful to introduce the inverse Fourier transform of $P_{\rm occ}(E)$, $P_{\rm occ}(\tau)$, via
\begin{eqnarray}
	P_{\rm occ}(E) = \int_{-\infty}^{+\infty} d \tau e^{i E \tau} P_{\rm occ}(\tau).
\end{eqnarray}
For a closed cavity, the kernel $P_{\rm occ}(\tau)$ is directly related to the characteristic function $\chi$ of the Glauber-Sudarshan $P$ function $\PP(\alpha)$, 
c.f.~Eqs.(\ref{eqs:PEFormula})-(\ref{eq:chi}) of the main text.  Recall that $\PP(\alpha)$ allows one to represent a given cavity state in phase space, c.f. Eq.~(\ref{eq:WFunction}) of the main text.

\subsection{Thermal state}

Consider a closed cavity in a thermal state having an average photon occupancy  $\bar n_{\rm th}=(e^{\hbar\Omega/k_BT}-1)^{-1} $.  Using Eqs.~(\ref{eqs:PEFormula}),(\ref{eq:chi}) of the main text, one finds
\beq
	P_{\rm occ}(\tau) = 
		\exp \left[  2\rho \bar{n}_{\rm th} (\cos(\Omega \tau)-1) \right].
	\label{eq:PtauTherm}
\eeq
The quasiprobability distribution $\PEph(E)$ is then
\begin{eqnarray}
	\PEph(E) =   \ee{-2 \rho \bar{n}_{\rm th} } \sum_{n,m =0}^{+\infty}
		\frac{(\rho \bar{n}_{\rm th})^{n+m}}{n!\, m!}  \delta\left(E - (n-m)\hbar \Omega \right)
		\equiv \sum_{n=-\infty}^{+\infty} p_{\rm occ, th}[n, \bar{n}_{\rm th}] \delta(E - n \hbar \Omega).
		\label{eq:pthDefn}
\end{eqnarray}
$\PEph(E)$  is just the convolution of two Poisson distributions: the first describes the Poisson absorption of photons at rate $\rho \bar{n}_{th}$, the second the 
Poissonian emission of photons at $\rho \bar{n}_{\rm th}$.  It thus has the form of a Skellam distribution.  Convolving in the vacuum distribution $P_0(E)$ given in Eq.~(\ref{eq:P0closed}) of the main text, one obtains the final distribution $P_{\rm tot}(E)$.  This continues to have the form of a Skellam distribution, and recovers
the expression for a thermal cavity which is well known from the standard theory of DCB \cite{IN}.

%

\subsection{Squeezed states}
Consider first the case where a cavity is prepared in a pure squeezed state which evolves without dissipation.  The initial cavity state is parametrized as:
\begin{equation}
	| r, \theta \rangle = 
		\exp \left( r e^{i \theta} \hat{a}^\dagger \hat{a}^\dagger - h.c. \right) | 0 \rangle,
\end{equation}
where $r$ is the squeeze parameter and the angle $\theta$ determines the orientation of the squeezed cavity quadrature.  In the absence of dissipation, the $P_{\rm occ}(E)$ function describing a cavity squeezed state can easily be calculated from characteristic function of this state.  Before averaging over the observation time $t$, we have:
\begin{equation}
	P_{\rm occ}(\tau,t) = 
	\exp \left[
		-4 \rho  \sinh^2 (r) \sin^2 \left(  \Omega \tau / 2 \right) 
	\right]
	\exp \left[
		2 \rho  \sinh (2r) \sin ^2\left(  \Omega \tau / 2 \right) ( \cos (2  \Omega t + \theta ))
	\right].
\end{equation}
$P_{\rm occ}(\tau)$ is obtained by averaging over $t$.
Without loss of generality, we shift the zero of time to absorb the phase $\theta$.  It is useful to first Fourier transform in the relative time variable $\tau$, but keep the dependence on observation time $t$.
Note that each factor above (for fixed $t$) has the same functional dependence on $\tau$ as the $P_{\rm occ}(\tau)$ for a thermal state (c.f. Eq.~(\ref{eq:PtauTherm})).    One thus obtains a simple convolution of two thermal distributions:
\begin{eqnarray}
	P_{\rm occ}(E,t) & = &
		\sum_n p_{\rm occ}[n, t] \delta(E - n \hbar \Omega), \\
	p_{\rm occ}[n,t] & = & \sum_{m} 
		p_{\rm occ,th}[n-m, \bar{n}_{\rm th} = \sinh^2{r} ] \cdot
		p_{\rm occ,th}[m, \bar{n}_{\rm th} = -(\sinh{2r} /2)  \cos (2  \Omega t )],\label{eq:S29}
\end{eqnarray}
where the weights $p_{\rm occ,th}[m, \bar{n}_{\rm th}]$ for a thermal distribution are defined in Eq.~(\ref{eq:pthDefn}).
Note that the second thermal distribution in the convolution has an effective temperature which is time dependent and {\it which can be negative} (i.e. for times where $\cos(2 \Omega t) > 0$).
This leads to negativity in $p_{\rm occ}[n,t]$, negativity which can persist even after averaging over the observation time $t$.  It thus is the origin of 
negativity in $P_{\rm occ}(E)$ for a squeezed state.  Further, note that when we average over $t$, even and odd photon number processes generated by the second thermal distribution will be impacted differently (as for small $\rho$, $p_{\rm occ, th}[m,\bar{n}_{\rm th}] \propto  (\bar{n}_{\rm th})^m$).  Thus, the above form also suggests the origin of the even-odd asymmetry in $P_{\rm occ}(E)$ for a cavity squeezed state.

If we time average $P_{\rm occ}(\tau,t)$, we find:
\begin{equation}
	P_{\rm occ}(\tau) = 
		e^{-4 \rho  \sinh ^2(r) \sin ^2\left(\frac{  \Omega \tau}{2}\right)} 
		I_0\left[2 \rho  \sinh (2r) \sin^2\left(\frac{  \Omega \tau }{2}\right) \right].
\end{equation}
Note that for a highly squeezed state, it is tempting to take the $r \rightarrow \infty$ limit, and make the approximations $\sinh^2 r \sim e^{2r}/4$ and $\sinh(2r) \sim e^{2r}/2$.  In this case, we could introduce $\rho_{\rm eff} = \rho e^{2r}/2$, and write:
\begin{equation}
	P_{\rm occ}(\tau) \simeq 
		e^{-\rho_{\rm eff}   \sin ^2\left(\frac{ \Omega \tau  }{2}\right)} 
		I_0\left[\rho_{\rm eff}  \sin^2\left(\frac{  \Omega \tau }{2}\right) \right].
\end{equation}
One can confirm that in this limit, $P_{\rm occ}(E)$ never exhibits any negativity.  It follows that even for very large squeeze parameters $r$, the presence of negativity depends crucially on the magnitude of $\rho$ and requires $\rho \sim 1$.  

We can also calculate $P_{\rm occ}(E)$ for an open (i.e.~damped) cavity that is prepared in a squeezed state by continuous driving:  squeezed input noise is continuously fed into the cavity.  In this case, the cavity state is Gaussian, and $P_{\rm occ}(E)$ is determined by the two-point phase phase correlator; this is easily calculated using standard Heisenberg-Langevin equations.  One finds
\begin{eqnarray}
	P_{\rm occ}(\tau, t) & = &
		\exp \left[  \rho \left( \Lambda_1 +  e^{-\kappa |\tau|/2} \Lambda_2 \right) \right],
\end{eqnarray}
where:
\begin{eqnarray}
	\Lambda_1 & = & 			
			-2  \sinh^2(r) - \sinh (2r) \cos (  \Omega \tau ) \cos (2  \Omega t ), \\
	\Lambda_2 & = &
			2   \sinh ^2(r) \cos ( \Omega \tau )+  \sinh (2 r) \cos (2  \Omega t ).	
\end{eqnarray}		
Note that despite the non-zero dissipation, we have undamped oscillations as a function of $\tau$, corresponding to processes where the photon energy is precisely $\Omega$.  	

\subsection{Fock states}

We denote the $P_{\rm occ}(\tau,t)$ function for an $n$-photon Fock state as 
$P_{{\rm occ}, n}(\tau,t)$.  It is given by:
\begin{eqnarray}
&P_{{\rm occ},n}(\tau,t) &=\langle n|  \ee{\lambda \acd}\ee{-\bar\lambda \ac}\ket{n}
=\sum_{p=0}^{+\infty} \frac{1}{p!^2} \langle n|(-|\lambda|^2\acd\ac)^p |n\rangle,\\
&&=\sum_{p = 0}^{+\infty} \frac{1}{p!}\binom n p \left(\ee{\ii\Omega \tau}-1+\ee{-\ii\Omega \tau}-1\right)^p,
\label{eq:Laguerre}
\end{eqnarray}
where we introduced $\lambda(\tau,t)=\sqrt{\rho}(\ee{i\Omega (t+\tau)}-\ee{i\Omega (t-\tau)})$. Note that the characteristic function does not depend on the observation time $t$, implying that there will be no average ac current across the tunnel junction.  Physically, this is a consequence of the rotational invariance of the Fock state phase space distribution.
Let us now discuss the above expression for a few specific choices of $n$.

\subsubsection{$n=1$ Fock state}
For $n=1$, Eq.~(\ref{eq:Laguerre}) becomes:
\begin{eqnarray}
&P_{{\rm occ}, 1}(\tau)&=1+\rho\left(\ee{\ii \Omega \tau}+\ee{-\ii \Omega \tau}-2\right).
\end{eqnarray}
The probability $P_{\rm occ}(E)$ is just given by the Fourier transform of $P_{\rm occ}(\tau)$ and reads 
\beq P_{{\rm occ}, 1}(E)=(1-2\rho)\delta(E)+\rho \left[  \delta(E+\hbar\Omega)+\delta(E-\hbar\Omega) \right].\eeq
The total $P(E)$ function including vacuum noise, $P_{\rm tot}(E)$ is given by a simple convolution with $P_0(E)$.  In the time domain, we obtain:
\begin{eqnarray}
&P_{{\rm tot},1}(\tau)	&=
	\ee{-\rho}\rho\ee{\ii\Omega \tau}+\ee{-\rho}\left(\rho-1\right)^2+\sum_{k=1}^{+\infty} \frac{\ee{-\rho}\rho^k}{(k+1)!}\left(\rho-(k+1)\right)^2\ee{-\ii k \Omega \tau}=\sum_{k=-\infty}^{+\infty} \frac{\ee{-\rho}\rho^k}{(k+1)!}\left(L_1^{(k)}(\rho)\right)^2\ee{-\ii k\Omega \tau},
\end{eqnarray}
where the  $L_n^{(k)}(\rho)$ are the generalized Laguerre polynomials. Negative probabilities have disappeared in this expression, as expected.

\subsubsection{$n=2$ Fock state}

For $n=2$, Eq.~(\ref{eq:Laguerre}) becomes:
\begin{eqnarray}
&P_{{\rm occ}, 2}(\tau)&=1+2\rho\left(\ee{\ii \Omega \tau}+\ee{-\ii \Omega \tau} -2 \right)+\frac{\rho^2}{2}\left(6+\ee{2\ii \Omega \tau}+\ee{-2\ii \Omega \tau} -4\left(\ee{\ii \Omega \tau}+\ee{-\ii \Omega \tau}\right) \right)\\
&&=(1-4\rho+3\rho^2)+2\rho(1-\rho)\left(\ee{\ii \Omega \tau}+\ee{-\ii \Omega \tau} \right)+\frac{1}{2}\rho^2\left( \ee{2\ii \Omega \tau}+\ee{-2\ii \Omega \tau} \right).
\end{eqnarray}
Again, negative probabilities for both zero and one photon absorption and emission processes are possible.  

Convolving in the vacuum absorption distribution $P_0(E)$ and remaining in the time domain, the full $P(E)$ function is given by:
\begin{eqnarray}
P_{{\rm tot},2}(\tau)&=\frac{\rho^2}{2}\ee{2\ii\Omega \tau}+\left[\frac{\rho^3}{2}+2\rho(1-\rho)\right]\ee{\ii\Omega \tau}+\left[\frac{\rho^4}{4}+2\rho^2(1-\rho)+(1-4\rho+3\rho^2)\right]\nonumber\\
&+\left[\frac{\rho^{5}}{2!3!}+2\rho(1-\rho)\left(\frac{\rho^2}{2}+1\right)+(1-4\rho+3\rho^2)\rho\right]\ee{-\ii\Omega \tau}  \nonumber \\
&+\sum_{k\geq 2} \left[\frac{\rho^{2}}{2}\left(\frac{\rho^{k+2}}{(k+2)!}+\frac{\rho^{k-2}}{(k-2)!}\right)+2\rho(1-\rho)\left(\frac{\rho^{k+1}}{(k+1)!}+\frac{\rho^{k-1}}{(k-1)!}\right)+(1-4\rho+3\rho^2)\frac{\rho^k}{k!}\right]\ee{-k\ii\Omega \tau}.
\end{eqnarray}
All these terms factorise nicely, and we obtain the simple closed expression for $P_{\rm occ}(\tau)$:
\begin{equation}
P_{{\rm tot},2}(\tau)=\ee{-\rho} \sum_{k=-\infty}^{+\infty} \frac{2\rho^k}{(k+2)!}
\left[ L_2^{(k)} (\rho) \right]^2 \ee{-k\ii\Omega \tau}.
\end{equation}

\subsubsection{$n>2$ Fock states}
In order to obtain a closed expression for $P_{\rm tot}(\tau)$ for an arbitrary Fock state, we first make use of the following property of Laguerre polynomials: 
\begin{equation}\label{eq:recursive}
 L_{n+1}(x)=\frac{1}{n+1}\left((2n+1-x)L_{n}(x)-nL_{n-1}(x)\right).
\end{equation}
This allows us to write the following recurrence relation between the $P_{{\rm tot},n}(E)$ function for different Fock states :
\beq\label{eq:recur}
P_{{\rm tot},n+1}(\tau)=\frac{1}{n+1}\left[\left(2(n-\rho)+1+\rho\left(\ee{\ii\Omega \tau}+\ee{-\ii\Omega \tau}\right)\right)
P_{{\rm tot},n}(\tau)-nP_{{\rm tot},n-1}(\tau)\right].
\eeq
We can expand $P_{{\rm tot},n}(\tau)$ in the usual manner in terms of weights for $k$-photon absorption/emission processes, using the fact that the maximum
number of photons that can be emitted is $n$:
\begin{equation}\label{eq:expansion}
	P_{{\rm tot},n}(\tau)=\sum_{k=-n}^{+\infty}p_{{\rm tot},n}[k] \ee{-\ii k\Omega \tau}.
\end{equation}
Inserting this form into Eq. (\ref{eq:recur}), we now have the following recurrence relations:
\begin{eqnarray}
&p_{{\rm tot},n+1}[-n-1]&=\frac{\rho}{n+1} p_{{\rm tot},n}[-n],\\
&p_{{\rm tot},n+1}[-n]&= \frac{1}{n+1}\left( \rho p_{{\rm tot},n}[-n+1]+\left(2(n-\rho)+1\right)p_{{\rm tot},n}[-n] \right),\\
&p_{{\rm tot},n+1}[k>-n]&=\frac{1}{n+1}\left( \rho(p_{{\rm tot},n}[k+1]
+p_{{\rm tot},n}[k-1])+ (2(n-\rho)+1)p_{{\rm tot},n}[k] -n p_{{\rm tot},n-1}[k] \right).
\end{eqnarray}

Motivated by the case $n=1$ and $n=2$, one finds by inspection that the following compact expression
\beq
	p_{{\rm tot},n}[k]=\frac{\ee{-\rho}\rho^kn!}{(k+n)!} \left[L_n^{(k)}(\rho)\right]^2,
	\label{eq:FockWeightSI}
\eeq
safisfies the above recursion relations.  This is Eq.~(\ref{eq:FockWeight}) in the main text.   As one can explicitly verify that this solution is correct for $n=1$ and $n=2$, it is thus necessarily unique.
The proof is  straightforward provided we use appropriate relations between the Laguerre polynomials such as:
\begin{eqnarray}
	&L_n^{(p)}(\rho) & = L_n^{(p+1)}(\rho) - L_{n-1}^{(p+1)}(\rho) \\[10pt]
	&n L_n^{(p)}(\rho) & = (n + p )L_{n-1}^{(p)}(\rho) - \rho L_{n-1}^{(p+1)}(\rho), \\[10pt]
	&n L_n^{(p+1)}(\rho) & =(n-\rho) L_{n-1}^{(p+1)}(\rho) + (n+p)L_{n-1}^{(p)}(\rho) \\[10pt]
	&\rho L_n^{(p+1)}(\rho) & = (n+p)L_{n-1}^{(p)}(\rho)-(n-\rho)L_n^{(p)}(\rho)\\
	&L_n^{(p)}(\rho)	&=  \frac{p+1-\rho}n  L_{n-1}^{(p+1)}(\rho)- \frac \rho n L_{n-2}^{(p+2)}(\rho). 
\end{eqnarray}

Eq.~(\ref{eq:FockWeight}) shows explicitly that the total probability (i.e.~including the contribution of vacuum fluctuations) to 
absorb or emit any given number of photons remain positive, regardless of the Fock state; this matches our general result. 
Nonetheless, it remains possible that the weight $p_{{\rm tot},n}[k]$ can be exactly zero for $k > -n$; this requires in general a careful tuning of the parameter $\rho$.  Such a cancellation would be {\it impossible} if $p_{{\rm occ},n}[j]$ were
all positive.  Hence, it serves as direct proof that we have a quantum state in the cavity.
In particular, consider $p_{{\rm tot},n}[0]$, the probability that an electron tunnels without any energy exchange with the cavity.  This quantity is directly
given by the excess current noise at zero frequency, see Eq.~(\ref{eq:DeltaS}) in the main text.  From the above expressions, we see that $p_{{\rm tot},n}[0]$
can be made zero if $\rho$ is 
tuned to be a root of the $n^{\rm th}$ Laguerre polynomial. The smallest root of the $n^{\rm th}$ Laguerre polynomial goes as $\sqrt{2}/n$, suggesting that the bigger the photon number of the cavity Fock state, the lower the minimum value of $\rho$ needed to see evidence of negativity.
 	


\section{Effects of finite temperature}


In the main text, we focused primarily on the regime of zero temperature for both the cavity and the electronic conductor.  We now consider the effects of non-zero temperature (both for the cavity state, and for the electrons in the tunnel junction leads).  We consider the realistic case of a microwave cavity with frequency $\Omega / 2 \pi$ in the range from $5 - 10$ GHz and we assume its temperature is maintained between $15$ and $30$ mK; this implies $\beta\hbar\Omega\in [15,30]$.  For such low temperatures, we find the modification of $P_{\rm tot}(E)$ compared to the zero temperature case is negligible; the main effect of temperature is thus through the tunnelling rates $\Gamma(E)$, namely the smearing of the sharp Fermi distribution in the metallic leads.  The net result
is that the heights of the various plateaus in the differential conductance are not affected by temperature but the transitions between them are rounded off, as shown in Fig.3.  This
is seen in Fig.~(\ref{fig:FiniteTemperatures}).     

%
%

\begin{figure*}[htpb]
	\begin{center}
		\includegraphics[width=0.4\textwidth]{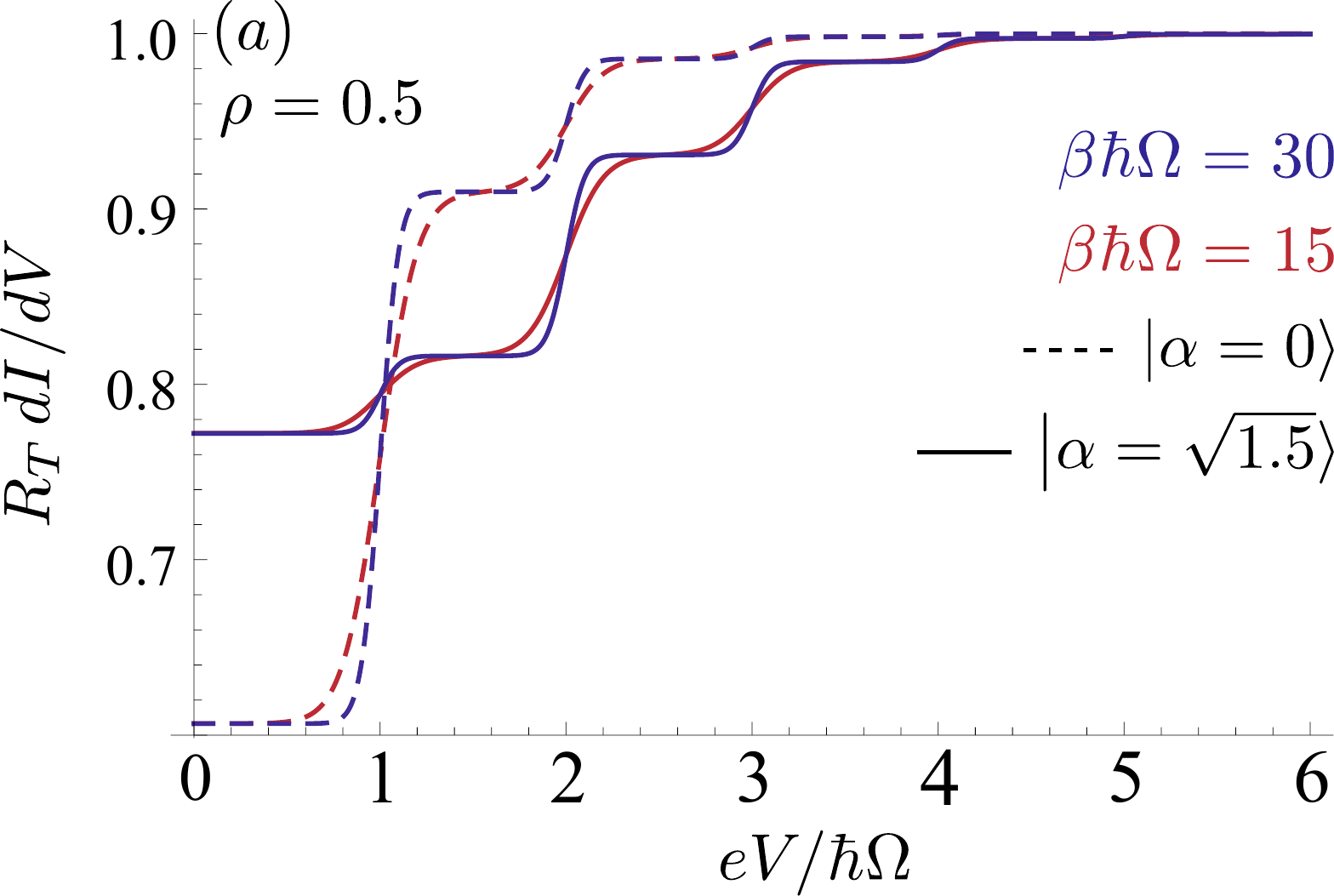}
		\hspace{1cm}
		\includegraphics[width=0.4\textwidth]{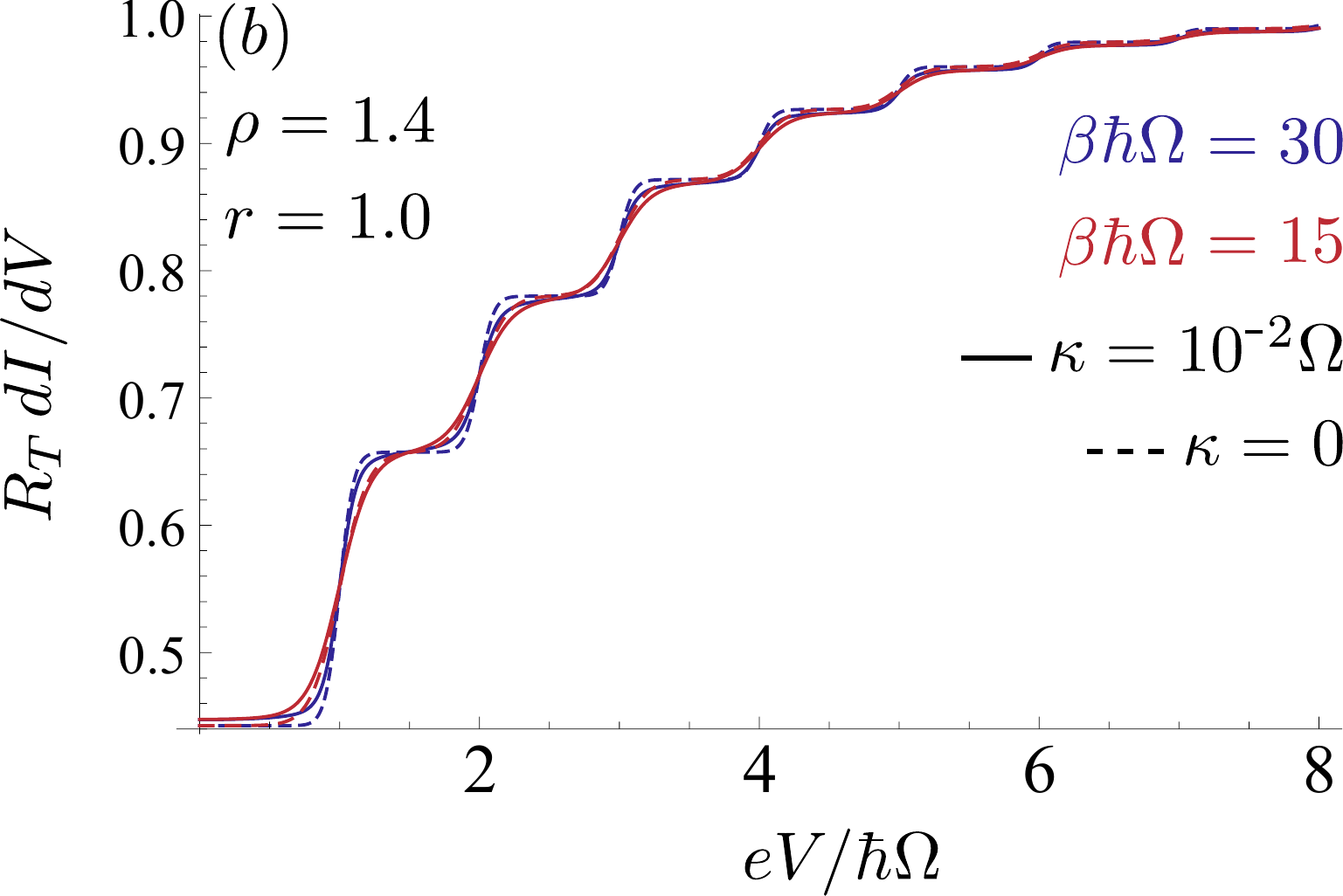}
		\caption{
			\label{fig:FiniteTemperatures}
		(a) Effects of the temperature on the differential conductance of a tunnel junction coupled to a ground-state cavity (dashed lines) and to a cavity in a coherent state of amplitude $\left|\alpha=\sqrt{1.5}\rangle\right.$ The heights of the plateaus are not affected by the temperature, but the smearing of transitions decreases their effective width. (b) As for part (a), but for a cavity prepared in a squeezed state with (solide line) and without damping (dashed line).  For these parameters, the smearing of
		transitions between conductance plateaus is more affected by the non-zero cavity damping $\kappa$ than by temperature.
		}
	\end{center}
\end{figure*}


\section{Using positivity of $P_{\rm occ}(E)$ to bound transport measurements}

As discussed in the main text and in the methods section, using the fact that $p_{0}[k]$ is a known distribution (Poissonian with mean $\rho$) one can obtain inequalities for both the heights of the differential conductance plateaus and for the weights $p_{\rm tot}[k]$ that {\it must} be satisfied for any cavity state where the quasi-probability
distribution $p_{\rm occ}[k]$ is positive definite.  As discussed, this will fail to be true when the cavity is prepared in a truly non-classical state, and hence these bounds may be violated by such states.  Three such bounds are used in the main text and two of these are given explicitly in the methods section.  Here, we derive them all.

\subsection{Bound on height of second conductance plateau}

We first obtain the bound on the height of the second conductance plateau $h_2$ given in Eq.~(\ref{eq:h2inequality}) of the main text and used in Fig.~\ref{fig:FockStateConductance}. This bound is useful in inferring the non-classicality of the Fock state via a transport measurement. Using Eq.~(\ref{eq:hn}) from the main text, expressing $p_{\rm tot}[k]$ as a convolution [c.f. Eqs.~(\ref{eq:PEclosed})-(\ref{eq:PEweights}) of the main text], and then using the evenness of $p_{\rm occ}[k]$, we have
\begin{eqnarray}
h_2-h_1 &= & p_{\rm tot}[+1] - p_{\rm tot}[-1] = p_0[1] p_{\textrm{occ}}[0]+p_{0}[2]p_{\textrm{occ}}[1]-\sum^{+\infty}_{k = 2} \left(p_{0}[k-1]-p_{0}[k+1]\right) p_{\textrm{occ}}[k] .  \label{eq:h2minh1a} 
\end{eqnarray}
For $k \geq 2$ and $\rho^2 < k (k + 1) $ we have $p_{0}[1] > p_{0}[k-1]-p_{0}[k+1] > 0 $. Therefore, for any $p_{occ}[k]$ distribution that is positive definite, we have
\begin{eqnarray}
h_2-h_1> p_0[1] p_{\textrm{occ}}[0]+p_{0}[2]p_{\textrm{occ}}[1]-p_{0}[1] \sum^{+\infty}_{k = 2} p_{\textrm{occ}}[k] = -\frac 12 p_0[1]+\frac 32 p_0[1] p_{\occ}[0] + \left(p_0[1]+p_0[2]\right) p_{\occ}[1]. \label{eq:etape1}  
\end{eqnarray}
The equality in Eq.~(\ref{eq:etape1}) follows from the normalisation of $p_{\rm occ}[k]$, $ 2\sum^{+\infty}_{k = 2}p_{\occ} [k] = 1-p_{\occ}[0]-2p_{\occ}[1] $.

The $p_{\rm occ}[k]$ weights in Eq.~(\ref{eq:etape1}) are unknown, so we wish to bound them. Again using Eq.~(\ref{eq:hn}) from the main text, we find
\begin{eqnarray}
h_3-h_2 - p_0[4]p_{\textrm{occ}}[2]& = & p_0[2] p_{\textrm{occ}}[0]+\left(p_0[1]+p_0[3]\right)p_{\textrm{occ}}[1]-\sum^{+\infty}_{k = 2}\left(p_0[k-2]-p_0[k+2]\right) p_{\textrm{occ}}[k] . \label{eq:h3minh2}
\end{eqnarray}
For $k \geq 2$ and $\rho^4 < (k+2)(k+1)k (k - 1)$ and $k_{\rm min} = 2$ we have $p_0[k-2] - p_0[k+2] > 0$ such that 
\begin{eqnarray}
h_3-h_2 -p_0[4]p_{\textrm{occ}}[2] < p_0[2] p_{\textrm{occ}}[0]+\left(p_0[1]+p_0[3]\right)p_{\textrm{occ}}[1] < \frac{3}{2} p_0[1] p_{\textrm{occ}}[0]+\left(p_0[1]+p_0[2]\right)p_{\textrm{occ}}[1] ,\label{eq:lastinequality}
\end{eqnarray}
where the second inequality holds for $\rho^2 < 3$. Now using the transitive property of the inequality and the fact that $p_{\rm occ}[k] \leq 1$, Eqs.~(\ref{eq:etape1}) and (\ref{eq:lastinequality}) lead to Eq.~(\ref{eq:h2inequality}) of the main text, a lower bound on $h_2$ for $\rho^2 < 3$ with a positive definite $p_{\rm occ}[k]$. 

\subsection{Bound on $p_{\tot}[n]$}

We next turn to the bound on $p_{\tot}[n]$ stated in Eq.~(\ref{eq:pmin}) of the main text, one that can be violated by a squeezed state in the cavity, as shown in Fig.~\label{fig:SqueezedState}(a) of the main text.  
Again, we bound the behaviour of $p_{\tot}[k]$ assuming only that $p_{\occ}[k]$ is positive definite. We start by using the fact that $p_{\tot}[n]$ is a convolution between $p_0[k]$ and $p_{\rm occ}[k]$,
\begin{subequations}
\begin{eqnarray}
p_{\rm tot} [n] = \sum^{+\infty}_{k = 0} p_0 [k] p_{\rm occ}[n-k ] & = & p_0 [0] p_{\rm occ}[n ] + \frac{ p_0[ 1 ] }{ p_0[0] } p_0 [0] p_{\rm occ}[ n-1 ]  + \sum^{+\infty}_{k = 2} \frac{ p_0[ k ] }{ p_0[0] } p_0 [0] p_{\rm occ}[ n-k ] , \label{eq:ptotExpanded1} \\
p_{\rm tot} [n-1 ] - p_{\rm tot}[n-2] & = & p_0 [0] p_{\rm occ}[n-1] + \sum^{+\infty}_{k = 2} \left( p_0[k-1] - p_0[k-2] \right) p_{\rm occ}[n-k ] . \label{eq:weightsdifference}
\end{eqnarray}
\end{subequations}
Substituting Eq.~(\ref{eq:weightsdifference}) into Eq.~(\ref{eq:ptotExpanded1}) we find
\begin{equation}
p_{\rm tot} [n ] = p_0[0] p_{\rm occ}[n] + \frac{ p_0[1] }{ p_0[0] } \left[ p_{\rm tot}[n-1] - p_{\rm tot}[n-2] + \sum^{+\infty}_{k = 2} \left( \frac{ p_0[0] p_0[k] }{ p_0[1] } - p_0[k-1] + p_0[k-2] \right) p_{\rm occ}[n-k] \right] . \label{eq:convwithsub}
\end{equation}
Now the term in braces in Eq.~(\ref{eq:convwithsub}) is simply $ e^{-\rho}  k! \rho^{k-2} (k-\rho )(k-1)/k! $, which is greater than zero for $k \geq 2$ and $\rho < 2$. With $p_{\rm occ}[k]$ positive definite we then have Eq.~(\ref{eq:pmin}) of the main text, a lower bound for $p_{\tot}[n]$ for $\rho < 2$. 
	
\subsection{Bound on height of fourth conductance plateau}

Now we will derive a bound on the height of the fourth conductance plateau, a bound that is useful for inferring the non-classicality of a squeezed state directly from a transport measurement, as shown in Fig.~\ref{fig:SqueezedState}(b) of the main text. The calculations described below follow from no more than Eqs.~(\ref{eq:h1}) and (\ref{eq:hn}) of the main text, and the assumption of a positive definite $p_{\rm occ}[k]$. This assumption shall be made throughout the calculation, but henceforth, will not be explicitly stated when it is employed. 

\subsubsection{Height differences of conductance plateaus}

We start by expressing the height differences of the conductance plateaus in terms of the weights of the $P(E)$ functions. Using Eq.~(\ref{eq:hn}) of the main text we can express the difference between the heights of the second and third conductance plateaus, as in Eq.~(\ref{eq:h3minh2}),
\begin{eqnarray}
h_3 - h_2 & = & p_0[2] p_{\rm occ}[0] + (p_0[1] + p_0[3]) p_{\rm occ}[1] + p_0[4] p_{\rm occ}[2] + \sum^{+\infty}_{k=3} \left( p_0[k+2] - p_0[k-2] \right) p_{\rm occ}[k] . \label{eq:first}
\end{eqnarray}
Subtracting $ \rho \left( 2p_0[2] + p_0[4] \right) p_{\rm occ}[1]/2$ from both sides of Eq.~(\ref{eq:first}) leads to 
\begin{eqnarray}
h_3 - h_2 - p_0[2]p_{\rm occ}[0] - \frac{\rho}{2} \left( 2p_0[2] + p_0[4] \right) p_{\rm occ}[1] - p_0[4] p_{\rm occ}[2] + \sum^{+\infty}_{k=3} \left( p_0[k-2] - p_0[k+2] \right) p_{\rm occ}[k] \nonumber \\
= \left[ p_0[1] + p_0[3] -\frac{\rho}{2} \left( 2p_0[2] + p_0[4] \right) \right] p_{\rm occ}[1] .
\end{eqnarray}
The right-hand-side is positive provided that $\rho < 1.525$, and consequently we have that the left-hand-side is greater than zero. Therefore, we can write
\begin{eqnarray}
\sum^{+\infty}_{k=3} \left( p_0[k-2] - p_0[k+2] \right) p_{\rm occ}[k] & > & - h_3 + h_2 + p_0[2]p_{\rm occ}[0] + \frac{\rho}{2} \left( 2p_0[2] + p_0[4] \right) p_{\rm occ}[1] + p_0[4] p_{\rm occ}[2] . \label{eq:h2h3Ineq}
\end{eqnarray} 

Again using Eq.~(\ref{eq:hn}) of the main text, we can also write an expression for the height difference of the third and fourth conductance plateaus, 
\begin{eqnarray}
\sum^{+\infty}_{k=4} \left( p_0[k-3] - p_0[k+3] \right) p_{\rm occ}[k] & = & - h_4 + h_3 + p_0[3] p_{\rm occ}[0] + (p_0[2] + p_0[4])p_{\rm occ}[1] + ( p_0[1] + p_0[5])p_{\rm occ}[2] \nonumber \\
& & + p_0[6] p_{\rm occ}[3 ] . \label{eq:h3h4Init}
\end{eqnarray}
Now we have $\rho \left( p_0[k-3] - p_0[k+3] \right)/2 > p_0[k-2] - p_0[k+2]$ for $\rho < 3.742$. This allows us to rewrite Eq.~(\ref{eq:h3h4Init}) as an inequality, and separating out the first term in the summation we have
\begin{eqnarray}
\sum^{+\infty}_{k=3} \left( p_0[k-2] - p_0[k+2] \right) p_{\rm occ}[k] & < &  \frac{\rho}{2} \left[ p_0[3] p_{\rm occ}[0] + (p_0[2] + p_0[4]) p_{\rm occ}[1] + (p_0[1] + p_0[5]) p_{\rm occ}[2]  \right. \nonumber \\
& & \left. + p_0[6] p_{\rm occ}[3] - h_4 + h_3 \right] + (p_0[1] - p_0[5]) p_{\rm occ}[3] . \label{eq:h3h4Ineq}
\end{eqnarray}

In Eqs.~(\ref{eq:h2h3Ineq}) and (\ref{eq:h3h4Ineq}) we have obtained an upper and lower bound on the sum of $p_{\rm occ}[k]$. Applying the transitive property of the inequality between these equations allows us to eliminate this sum, leaving us with
\begin{eqnarray}
\frac{\rho}{2} p_0[2] p_{\rm occ}[1] & < & h_3 - h_2 - \frac{\rho}{2} (h_4 - h_3) + \left( \frac{\rho}{2} p_0[3]  - p_0[2] \right) p_{\rm occ}[0] + \left[ \frac{\rho}{2} ( p_0[1] + p_0[5] ) - p_0[4] \right] p_{\rm occ}[2] \nonumber \\
& & + \left( p_0[1] - p_0[5] + \frac{\rho}{2} p_0[6] \right) p_{\rm occ}[3] . \label{eq:h2h3h4Ineq}
\end{eqnarray}
Eq.~(\ref{eq:h2h3h4Ineq}) is now an inequality relating the second, third and fourth conductance plateaus. 

Applying Eq.~(\ref{eq:hn}) of the main text to the first and second conductance plateaus, as was done for Eq.~(\ref{eq:h2minh1a}) yields  
\begin{eqnarray}
p_0[2] p_{\rm occ}[1] & = & h_2 - h_1 - p_0[1] p_{\rm occ}[0] + ( p_0[1] - p_0[3] ) p_{\rm occ}[2] + (p_0[2] - p_0[4]) p_{\rm occ}[3] \nonumber \\
& & + \sum^{+\infty}_{k=4} \left( p_0[k-1] - p_0[k+1] \right) p_{\rm occ}[k] . \label{eq:h1h2}
\end{eqnarray}
Dropping the summation in Eq.~(\ref{eq:h1h2}), on the grounds that the coefficients of $p_{\rm occ}[k]$ are positive for $\rho < 4.472$, we are left with an inequality. Subsequently multiplying both sides of the inequality by $\rho/2$ we find 
\begin{equation}
\frac{\rho}{2} p_0[2] p_{\rm occ}[1] > \frac{\rho}{2} (h_2 - h_1) - \frac{\rho}{2} p_0[1] p_{\rm occ}[0] + \frac{\rho}{2} \left( p_0[1] - p_0[3] \right) p_{\rm occ}[2] + \frac{\rho}{2} \left( p_0[2] - p_0[4] \right) p_{\rm occ}[3] . \label{eq:h1h2Ineq}
\end{equation}

Considering Eqs.~(\ref{eq:h2h3h4Ineq}) and (\ref{eq:h1h2Ineq}), we see that we have an upper and a lower bound on $p_{\rm occ}[1]$. Using the transitive property of the inequality we find
\begin{eqnarray}
h_3 - h_2 - \frac{\rho}{2} ( h_4 - h_3 + h_2 - h_1) & > & -\frac{\rho}{2} p_0[3] p_{\rm occ}[0] + \left[ p_0[4] - \frac{\rho}{2} (p_0[3] + p_0[5] ) \right] p_{\rm occ}[2] \nonumber \\
& & + \left[ - p_0 [1] + p_0[5] + \frac{\rho}{2} \left( p_0[2] - p_0[4] - p_0[6] \right) \right] p_{\rm occ}[3] . \label{eq:h1234}
\end{eqnarray}
We have now obtained a lower bound on a linear combination of heights of the first four conductance plateaus. However, the bound still contains the (unknown) $p_{\rm occ}[k] \, (k=0,2,3)$. Therefore, we now seek to bound these quantities in terms of the known $p_0[k]$ and the measurable $h_k \, (k=1,2,3,4)$. 

\subsubsection{Bounding $p_{\rm occ}[0]$}

Using Eq.~(\ref{eq:h1}) from the main text with the normalisation conditions $1 = \sum^{+\infty}_{n=-\infty} \sum^{+\infty}_{m=0} p_0[m] p_{\rm occ}[n-m]$ and $1 = \sum^{+\infty}_{m=0} p_0[m]$, we can show that
\begin{subequations}
\begin{eqnarray}
1 - h_1 & = & \sum^{+\infty}_{n=1} \sum^{+\infty}_{m=0} p_0[m] \left( p_{\rm occ}[n-m] - p_{\rm occ}[n+m]  \right) \\
& = & (1-p_0[0]) p_{\rm occ}[0] + 2 \sum^{+\infty}_{k=1} \left( 1 - \sum^{k-1}_{m=0} p_0[m] - p_0[k]/2 \right) p_{\rm occ}[k] . \label{eq:1minh1Sums}
\end{eqnarray} 
\end{subequations}
Truncating the summation on the right-hand-side of Eq.~(\ref{eq:1minh1Sums}) at the $p_{\rm occ}[3]$ term and rearranging leads to a bound on $p_{\rm occ}[0]$,
\begin{eqnarray}
p_{\rm occ}[0] & < & \frac{1-h_1}{1-p_0[0]} - 2 \frac{ 1 - p_0[0] - p_0[1]/2 }{ 1 - p_0[0] } p_{\rm occ}[1] - 2 \frac{1-p_0[0] - p_0[1] - p_0[2]/2}{ 1 - p_0[0] } p_{\rm occ}[2] \nonumber \\
& & - 2 \frac{ 1 - p_0[0] - p_0[1] - p_0[2] - p_0[3]/2 }{ 1 - p_0[0] } p_{\rm occ}[3] . \label{eq:p0bound}
\end{eqnarray}

Using the bound of Eq.~(\ref{eq:p0bound}) in Eq.~(\ref{eq:h1234}) leads to a coefficient on the $p_{\rm occ}[2]$ term (on the smaller side of the inequality), $ \rho p_0[3] (1-p_0[0] - p_0[1] - p_0[2]/2)/(1-p_0[0]) + p_0[4] - \rho \left( p_0[3] + p_0[5] \right)/2$. This coefficient is positive for $1.011 < \rho < 5.235$ and so the $p_{\rm occ}[2]$ term may be removed from the inequality. This leaves us with
\begin{eqnarray}
h_3 - h_2 - \frac{\rho}{2} \left( h_4 - h_3 + h_2 - h_1 \right) & > & -\frac{\rho}{2} p_0[3] \frac{1-h_1}{1-p_0[0]} + \rho p_0[3] \frac{1 - p_0[0] - p_0[1]/2}{1-p_0[0]} p_{\rm occ}[1] \nonumber\\
& & + \rho p_0[3] \frac{1 - p_0[0] - p_0[1] - p_0[2] - p_0[3]/2}{1-p_0[0]} p_{\rm occ}[3] \nonumber \\
& & + \left[ \frac{\rho}{2} \left( p_0[2] - p_0[4] - p_0[6] \right) - p_0[1] + p_0[5] \right] p_{\rm occ}[3] . \label{eq:useful}
\end{eqnarray}
Comparing Eq.~(\ref{eq:useful}) to Eq.~(\ref{eq:h1234}) we have one fewer unknown $p_{\rm occ}[k]$ in the bound; we now seek to reduce this number further. 

\subsubsection{Bounding $p_{\rm occ}[3]$}
Applying Eqs.~(\ref{eq:h1}) and (\ref{eq:hn}) of the main text iteratively we find the expressions for the height of the fourth conductance plateau, 
\begin{subequations}
\begin{eqnarray}
h_4 & = & p_{\rm tot}[0] + \sum^{3}_{k=1} \left( p_{\rm tot}[k] + p_{\rm tot}[-k] \right) + 2 \sum^{+\infty}_{k=4} p_{\rm tot}[-k] \\
& = & \sum^{+\infty}_{l=0} p_0[l] p_{\rm occ}[l] + \sum^{3}_{k=1} \sum^{+\infty}_{l=k} p_0[l-k] p_{\rm occ}[l] + \sum^{3}_{k=1} \sum^{+\infty}_{l=-k} p_0[l+k] p_{\rm occ}[l] + 2 \sum^{+\infty}_{k=4} \sum^{+\infty}_{l=k} p_0[l-k] p_{\rm occ}[l] . \label{eq:h4sums}
\end{eqnarray}
\end{subequations}
The second line follows from expanding the total $P(E)$ weights as convolutions of the vacuum and occupied $P(E)$ weights. Next we can write an expression for $1-h_4$ using Eq.~(\ref{eq:h4sums}) and the normalisation condition $1=\sum^{+\infty}_{m=0} p_0[m] p_{\rm occ}[0] + 2 \sum^{+\infty}_{m=0} p_0[m] \sum^{+\infty}_{k=1} p_{\rm occ}[k]$. Truncating the resulting expression at the $p_{\rm occ}[4]$ and $p_0[4]$ terms leads us to the inequality 
\begin{eqnarray}
1 - h_4 & > & \left( 1 - p_0[0] - p_0[1] - p_0[2] - p_0[3] \right) p_{\rm occ}[0] + \left( 1 - p_0[0] - p_0[1] - p_0[2] \right) p_{\rm occ}[1] \nonumber \\
& & + \left( 1 - p_0[0] - p_0[1] \right) p_{\rm occ}[2] + \left( 1 - p_0[0] \right) p_{\rm occ}[3] + \left( 1 - p_0[0] \right) p_{\rm occ}[4] . \label{eq:h4Comp}
\end{eqnarray}

Writing Eq.~(\ref{eq:h3h4Init}) as a lower bound for $h_3 - h_4$ by removing the summation, and adding the resulting inequality to that of Eq.~(\ref{eq:h4Comp}), we find
\begin{eqnarray}
1 + h_3 - 2h_4 & > & (1 - p_0[0] - p_0[6]) p_{\rm occ}[3] + (1-p_0[0] -2p_0[1] - p_0[5] ) p_{\rm occ}[2] \nonumber \\
& & + (1 - p_0[0] - p_0[1] - 2 p_0[2] - p_0[4] ) p_{\rm occ}[1] + (1 - p_0[0] - p_0[1] - p_0[2] - 2p_0[3] ) p_{\rm occ}[0] . \nonumber \\
& & \label{eq:h3h4Sum1}
\end{eqnarray}
The coefficient of $p_{\rm occ}[2]$ is positive for $\rho > 1.274$, and so we may remove it from this inequality. Further, $p_{\rm occ}[0] < (1-h_1)/(1-p_0[0])$, which follows from truncating Eq.~(\ref{eq:1minh1Sums}) at the first term. Using this bound, Eq.~(\ref{eq:h3h4Sum1}) can be written as a bound on $p_{\rm occ}[3]$,
\begin{eqnarray}
p_{\rm occ}[3] (1 - p_0[0] - p_0[6]) & < & 1 + h_3 - 2h_4 + (p_0[0] + p_0[1] + p_0[2] + 2p_0[3] - 1 ) \frac{1-h_1}{1-p_0[0]} \nonumber \\
& & + (p_0[0] + p_0[1] + 2 p_0[2] + p_0[4] - 1) p_{\rm occ}[1] . \label{eq:pocc3Bound}
\end{eqnarray}

\subsubsection{Final bounds on fourth conductance plateau}
Before stating our final bounds on the conductance plateaus we introduce some short-hand notation for the complicated functions of $\rho$ that arise:
\begin{subequations}
\begin{eqnarray}
\mathcal{A} & \equiv & \frac{- \rho (p_0[2] - p_0[4] + p_0[6])/2 + p_0[1] - p_0[5]}{ 1 - p_0[0] - p_0[6] } - \rho \frac{p_0[3]}{1-p_0[0]} \frac{1 - p_0[0] - p_0[1] - p_0[2] - p_0[3]/2 }{ 1 - p_0[0] - p_0[6] } , \\
\mathcal{B} & \equiv & 1/(1-p_0[0]) , \\
\mathcal{C} & \equiv & p_0[0] + p_0[1] + p_0[2] + 2p_0[3] - 1 .
\end{eqnarray}
\end{subequations}
Note that $\mathcal{A} > 0$ for $\rho < 1.859$, $\mathcal{C} > 0$ for $\rho < 2.318$, and $\mathcal{B}$ is unconditionally positive. 

Now using the bound of Eq.~(\ref{eq:pocc3Bound}) in Eq.~(\ref{eq:useful}) we see that the coefficient of $p_{\rm occ}[1]$ (appearing on the smaller side of the inequality), $ \rho \mathcal{B} p_0[3] (1 - p_0[0] - p_0[1]/2) + \mathcal{A} ( 1-p_0[0] - p_0[1] - 2p_0[2] - p_0[4] ) $, is positive for $1.042 < \rho < 5.212$, and therefore the $p_{\rm occ}[1]$ term may be discarded. This leaves us with an inequality which may be expressed as an upper bound on the height of the fourth conductance plateau, 
\begin{equation}
h_4 ( 2\mathcal{A} + \rho/2 ) < \mathcal{A} (1+ \mathcal{B} \mathcal{C} ) + \mathcal{B} \rho p_0[3]/2 - \left[ \mathcal{A} \mathcal{B} \mathcal{C} - \rho ( 1 - \mathcal{B} p_0[3] )/2 \right] h_1 - ( 1 + \rho/2 ) h_2 + (1 + \rho/2 + \mathcal{A} ) h_3 .
\end{equation}
This bound is valid for $1.274 < \rho < 1.525$. We stress that these bounds are expressed in terms of the known $p_0[k]$ and the measurable $h_k \, (k=1,2, 3)$. The bounds apply for a classical distribution in the coupled electromagnetic mode, and may be violated in the presence of a non-classical field.


\end{widetext}
\end{document}